\documentclass[namedreferences]{SolarPhysics}
\usepackage[optionalrh]{spr-sola-addons} 
\usepackage{graphicx}
\usepackage{color}
\usepackage{url}             

\newcommand{\adv}{    {\it Adv. Space Res.}}

\newcommand{\aap}{    {\it Astron. Astrophys.}}

\newcommand{\apj}{    {\it Astrophys. J.}}
\newcommand{\apjl}{   {\it Astrophys. J. Lett.}}

\newcommand{\jgr}{    {\it J. Geophys. Res.}}

\newcommand{\pasj}{   {\it Publ. Astron. Soc. Japan}}

\newcommand{\solphys}{{\it Solar Phys.}}

\newcommand{\ssr}{    {\it Space Sci. Rev.}}

\begin{document}
\begin{article}
\begin{opening}

\title{Relations between Microwave Bursts and near-Earth High-Energy
    Proton Enhancements and their Origin}

\author{V.V.~\surname{Grechnev}$^{1}$\sep
        V.I.~\surname{Kiselev}$^{1}$\sep
        N.S.~\surname{Meshalkina}$^{1}$\sep
        I.M.~\surname{Chertok}$^{2}$}

\runningauthor{Grechnev et al.} \runningtitle{Relations between
microwave bursts and high-energy protons}

\institute{$^{1}$ Institute of Solar-Terrestrial Physics SB RAS,
                  Lermontov St.\ 126A, Irkutsk 664033, Russia
                  email: \url{grechnev@iszf.irk.ru} email: \url{valentin_kiselev@iszf.irk.ru} email: \url{nata@iszf.irk.ru}\\
           $^{2}$ Pushkov Institute of Terrestrial Magnetism,
                  Ionosphere and Radio Wave Propagation (IZMIRAN), Troitsk, Moscow, 142190 Russia
                  email: \url{ichertok@izmiran.ru}}

\date{Received ; accepted }

\begin{abstract}
We further study the relations between parameters of bursts at 35
GHz recorded with the Nobeyama Radio Polarimeters during 25 years,
on the one hand, and solar proton events, on the other hand
(Grechnev et al. 2013: PASJ 65, SP1, S4). Here we address the
relations between the microwave fluences at 35 GHz and near-Earth
proton fluences above 100 MeV in order to find information on their
sources and evaluate their diagnostic potential. A correlation was
found to be pronouncedly higher between the microwave and proton
fluences than between their peak fluxes. This fact probably reflects
a dependence of the total number of protons on the duration of the
acceleration process. In events with strong flares, the correlation
coefficients of high-energy proton fluences with microwave and soft
X-ray fluences are higher than those with the speeds of coronal mass
ejections. The results indicate a statistically larger contribution
of flare processes to high-energy proton fluxes. Acceleration by
shock waves seems to be less important at high energies in events
associated with strong flares, although its contribution is probable
and possibly prevails in weaker events. The probability of a
detectable proton enhancement was found to directly depend on the
peak flux, duration, and fluence of the 35 GHz burst, while the role
of the Big Flare Syndrome might be overestimated previously.
Empirical diagnostic relations are proposed.

\end{abstract}

\end{opening}

\section{Introduction}
  \label{S-introduction}

The problems of the origin of solar proton events (SPEs) and their
diagnostics are hotly debated over almost half a century. Two
concepts of their origin are considered and even contrasted (see,
\textit{e.g.}, \opencite{Kallenrode2003}; \opencite{Grechnev2008};
\opencite{Aschwanden2012}; \opencite{Reames2013};
\opencite{Trottet2015} for a review and references). The
flare-acceleration concept relates the SPE sources to flare
processes in coronal magnetic fields of active regions, manifested
particularly in X-ray and microwave emissions. The
shock-acceleration concept relates the major SPE sources to bow
shocks driven by fast coronal mass ejections (CMEs).

There are convincing arguments in favor of both the flare and shock
origin of SPEs. Gamma-rays concurrent with other flare emissions
favor the hypothesis of acceleration of heavy particles in flares
simultaneously with electrons (see, \textit{e.g.},
\opencite{Chupp2009}; \opencite{Vilmer2011}). On the other hand,
\textit{in situ} measurements of the particle composition, such as
the iron charge state, Fe/O ratio, and others appear to favor the
shock-acceleration of ions at normal coronal temperatures (see,
\textit{e.g.}, \opencite {Reames2013}). Note that such measurements
are limited to moderate ion energies, while acceleration of heavier
ions is indeed more effective by Fermi mechanisms operating in
shock-acceleration. The apparent delay of the particle escape near
the Sun (\citeauthor{Reames2009}, \citeyear{Reames2009, Reames2013})
does not seem to be a reliable indication of their exceptional
shock-acceleration, because trapped flare-accelerated particles can
escape from closed coronal structures after their delayed
reconnection with open structures in the course of the CME expansion
\cite{Masson2012, Grechnev2013b}. It is also possible that the
alernative concepts are based on different observations subjected to
selection effects.

The sources of the particle acceleration in flares and by shock
waves are considered to be remote and independent of each other.
The concepts of their origin are mainly based on hypotheses
proposed in past decades, when opportunities to observe solar
phenomena were much poorer than now. The well-known fact of a
reduced proton productivity of short-duration events (referring to
the soft X-ray (SXR) emission) led to a hypothesis about
predominance of different acceleration mechanisms in `impulsive'
and `gradual' events (see, \textit{e.g.}, \opencite{Croom1971};
\opencite{Cliver1989}; \citeauthor{Reames2009},
\citeyear{Reames2009, Reames2013}; and references therein).

However, recent observational studies have revealed a closer
association between solar eruptions, flares, shock waves and CMEs,
than previously assumed. It was found that the CME acceleration
pulse occurs almost simultaneously with hard X-ray and microwave
bursts (\opencite{Zhang2001}; \citeauthor{Temmer2008},
\citeyear{Temmer2008, Temmer2010}). The helical component of the
CME's flux rope responsible for its acceleration is formed by
reconnection which also causes a flare \cite{Qiu2007}. A detailed
quantitative correspondence has been established between the
reconnected magnetic flux and the rate of the flare energy release
\cite{Miklenic2009}. Most likely, a shock wave is typically excited
by an erupting flux rope as an impulsive piston inside a developing
CME during the rise phase of the hard X-ray and microwave bursts
(\citeauthor{Grechnev2011}, \citeyear{Grechnev2011, Grechnev2013a}).
Then the shock wave detaches from the piston and quasi-freely
propagates afterwards like a decelerating blast wave. Its transition
to the bow-shock regime is possible later, if the CME is fast
\cite{Grechnev2015}. Thus, parameters of the CME and shock wave
should be related to those of a corresponding flare, and the
traditional contrasting of the acceleration in a flare and by a
shock might be exaggerated. Some aspects of the correspondence
between the parameters of flares, CMEs, shock waves, and SPEs have
been stated by \inlinecite{Nitta2003} and
\inlinecite{Gopalswamy2012}.

By taking account of recent results, one can expect a
correspondence between the parameters of SPEs and microwave
bursts. Indeed, the correlation between SPEs and strong
high-frequency radio bursts has been known for many decades
(\textit{e.g.}, \opencite{Croom1971};
\opencite{CastelliBarron1977}; \opencite{Akinian1978};
\opencite{Cliver1989}, \opencite{Melnikov1991}). Alternatively,
\inlinecite{Kahler1982}, advocating the shock-related origin of
SPEs, explained this association by the `Big Flare Syndrome'
(BFS), \textit{i.e.}, a general correspondence between the energy
release in an eruptive flare and its various manifestations. Thus,
different flare parameters should correlate with each other
regardless of any physical connection between them. The basic
concept is clear, while the measure of the degree of correlation
due to the BFS used by \inlinecite{Kahler1982} does not seem to be
obvious. Assuming a single source for accelerated protons and
heavier ions, he concluded that normal coronal temperatures of the
ions ruled out the flare-related origin of the protons. Thus, the
correlation with the thermal soft X-ray flux (1--8~\AA) was
considered as a measure of the BFS contribution. On the other
hand, the mentioned reasons indicate the origins of protons in
both flare-related and shock-related accelerators, whose
efficiency can be largely different for different particles. It is
difficult, if possible, to distinguish between different sources
of the protons. For these reasons, \inlinecite{Kahler1982} might
have somewhat overestimated the role of the BFS. A number of later
studies interpreting observational results in terms of traditional
hypotheses apparently supported the shock-acceleration concept
(\textit{e.g.}, \opencite{Tylka2005}; \opencite{Rouillard2012},
and others), that have led to an underestimation of diagnostic
opportunities of microwave bursts. Nevertheless, it seems worth to
analyze the relations between flare microwave bursts and SPEs,
irrespective of their origin.

These relations were considered previously in a number of studies,
but mostly at frequencies $ < 17$~GHz (see, \textit{e.g.},
\opencite{Akinian1978}; \opencite{Cliver1989}). These relatively low
frequencies can belong to either the optically thin or thick branch
of the gyrosynchrotron spectrum, causing the ambiguity of the
results and complicating their interpretation. This difficulty was
overcome by \inlinecite{Chertok2009} through measuring microwave
fluxes at two different frequencies.

Microwave emissions at higher frequencies in the optically thin
regime seem to be most sensitive to large numbers of high-energy
electrons gyrating in strong magnetic fields, being thus directly
related to the energy release rate in the flare--CME formation
process during its main (impulsive) phase. The frequency of~35 GHz
is the highest one, at which stable long-term observations are
available, thanks to the operation of the Nobeyama Radio
Polarimeters (NoRP; \opencite{Nakajima1985}). All of these
circumstances determined our choice of the analyzed data.

In our previous study \cite{Grechnev2013b}, we mainly analyzed the
relations between peak fluxes at 35~GHz, $F_{35} \geq 10^3$~sfu
(1~sfu~$ = 10^{-22}$~W~m$^{-2}$~Hz$^{-1}$), recorded with NoRP
since 1990 to 2012, on the one hand, and peak fluxes of SPEs
$>100$~MeV, $J_{100}$, on the other hand. Most events showed a
scattered direct tendency between the microwave and proton peak
fluxes. Considerable SPEs were revealed even after east solar
sources, if the microwave bursts were strong enough.

A better correspondence might exist between some combinations of
the time-integrals (fluences) of proton fluxes and microwave
bursts (see, \textit{e.g.}, \opencite{Kahler1982};
\opencite{Chertok1990}; \opencite{Trottet2015}). In this study we
consider these combinations.

An additional aspect of our analysis was inspired by a recent
study of \inlinecite{Trottet2015}, who analyzed the correlations
between proton fluxes in a range of 15--40 MeV and parameters
characterizing flares and CMEs. Their analysis revealed
significant correlations between the peak proton flux, on the one
hand, and the start-to-peak SXR fluence and CME speed, on the
other hand. Neither the microwave fluence nor the SXR peak flux
provided significant contribution to the total correlations. The
results indicate that both flare-accelerated and shock-accelerated
protons contribute to near-Earth fluxes in this energy range.
\inlinecite{Trottet2015} and \inlinecite{Dierckxsens2015} revealed
indications at the domination of shock-acceleration for protons
with energies below 10--20~MeV and flare-acceleration for higher
energies, but the statistical significance of this finding was
insufficient. Our data set allows us verifying this statement.

Our aim in this respect is not advocating either of the concepts
of the SPE origin. We try instead to understand how the results of
our analysis as well as those of different studies (sometimes
seemingly incompatible) could be to reconciled with each other to
form a probable consistent picture. In the course of our study, we
endeavor to find what the 35 GHz radio bursts can tell us about
SPEs, to reveal diagnostic opportunities of these radio bursts to
promptly estimate a probable importance of a forthcoming
high-energy SPE, and to highlight promising ways to further
investigate into the SPE problem.

A number of mentioned studies considered a sample of SPEs selected
by some criteria and analyzed parameters of responsible solar
eruptive events. Our reverse approach misses many SPEs after
weaker bursts, but it is natural for the diagnostic purposes and
promises understanding how parameters of microwave bursts are
related with the proton productivity of solar events.

Section~\ref{S-statistics} considers statistical relations between
the peak flux and duration of the 35 GHz burst and the probability
of a proton enhancement, as well as between the microwave and proton
fluences. Section~\ref{S-origin} analyzes correlations between
proton fluences and parameters characterizing flares and shock
waves, and examines which of these correlations are significant.
Section~\ref{S-discussion} discusses the results and presents the
main conclusions.


\section{Statistical Analysis of Parameters of Microwave Bursts and Proton Fluxes $> 100$~MeV}
 \label{S-statistics}

\subsection{Data}
 \label{S-data}

Data lists of microwave bursts recorded by NoRP are posted on the
Web site \url{http://solar.nro.nao.ac.jp/norp/html/event/}. We
considered all microwave bursts with peak flux densities at 35~GHz
$F_{35} \geq 10^3$~sfu. This criterion has selected 104 bursts. We
also looked for proton enhancements (\textit{e.g.},
\opencite{Kurt2004}) with peak fluxes $J_{100}> 10$~pfu (1~pfu = 1
particle~cm$^{-2}$~s$^{-1}$~sr$^{-1}$) not to miss big SPEs after
weaker microwave bursts, and revealed seven additional events.
Three of them were caused by backside sources, whose microwave
emission could not reach Earth. No conclusions can be drawn about
these events, and they were excluded from further analysis. Four
large SPEs occurred after moderate microwave bursts with $F_{35} <
10^3$~sfu. Two of them caused ground level enhancements of cosmic
ray intensity (GLEs): 2000-11-08, 2001-12-26 (GLE63), 2002-04-21,
and 2012-05-17 (GLE71).

Automatically processed digital NoRP data in the XDR (IDLsave)
format are accessible via
\url{ftp://solar-pub.nao.ac.jp/pub/nsro/norp/xdr/}. The technique
to accurately process NoRP data and to evaluate quantitative
parameters of the bursts is described in
\inlinecite{Grechnev2013b}. For each event we recalibrated the
pre-burst level, which was often not perfect. This constant level
was subtracted in calculating total microwave fluences. The
contribution of the thermal bremsstrahlung was estimated from SXR
GOES data for four mentioned proton-abundant events. It was 42\%
for the 2000-11-08 event, 31\% for 2012-05-17, 19\% for
2002-04-21, and 18\% for 2001-12-26. The thermal contribution to
the remaining stronger bursts with peak fluxes $F_{35} \geq
1000$~sfu was neglected.

Digital data of GOES proton monitors are available at
\url{http://satdat.ngdc.noaa.gov/sem/goes/data/new_avg/}. The total
proton fluences were calculated for the integral proton channel
$E_{p}> 100$~MeV for the whole time of a proton enhancement with
subtraction of a constant background level. If an SPE overlapped
with a decay of a preceding event, then the background was fit with
an exponential function.

The data on events with the analyzed microwave bursts, corresponding
proton enhancements, CMEs, and calculated parameters are presented
in Table~\ref{T-data_table}. The events are categorized according to
their peak fluxes at 35~GHz, $F_{35}$, similar to the GOES
classification. These are mX (microwave-eXtreme,  $F_{35} >
10^{4}$~sfu), mS (microwave-Strong, $10^{3}$~sfu $ < F_{35} <
10^{4}$~sfu), mM (microwave-Moderate, $10^{2}$~sfu $< F_{35} <
10^{3}$~sfu. The behind-the-limb events, whose microwave emission
could not be detected, are categorized as mO (microwave-Occulted)
events.

The list of events presented by \inlinecite{Grechnev2013b} was
supplemented with events since late 2012 to March 2015 and events
93 and 94, missing in the NoRP Event List. A number of typos have
been corrected. The table is supplemented with the calculated
total microwave and proton fluences, start-to-peak SXR fluences,
and the CME speeds, if known. They were taken from the on-line CME
catalog (\opencite{Yashiro2004};
\url{http://cdaw.gsfc.nasa.gov/CME_list/}) containing the
measurements from SOHO/LASCO data \cite{Brueckner1995}. An
atypical event 5 (SOL1991-05-18), previously assessed as a
non-SPE, was reconsidered. This long-duration mX event was
associated with an X2.8 flare, type IV and II bursts
(\textit{i.e.}, a CME and shock wave); thus, an SPE is expected in
any case. Unlike an apparently well-connected position (N32W87), a
related SPE had a long-lasting rise of more than half a day
\cite{Sladkova1998} typical of events with far-east sources.
\inlinecite{Chertok2009} assumed an unfavorable connection between
its source and Earth that is supported by the occurrence of a
geomagnetic storm on 17--19 May with Dst up to $-105$~nT
(\url{http://wdc.kugi.kyoto-u.ac.jp/dst_final/199105/index.html}).

 \begin{kaprotate}
 \begin{table*}
 \caption{Analyzed events}
\label{T-data_table}
 \begin{tabular*}{\maxfloatwidth}{rcccrcrrrrrrcr}
 \hline

No & Date & \multicolumn{4}{c}{Flare} & \multicolumn{3}{c}{Microwave burst} & \multicolumn{4}{c}{Protons near Earth} &$V_\mathrm{CME}$ \\

& & \multicolumn{1}{c}{$T_\mathrm{peak}$} &
\multicolumn{1}{c}{GOES} & \multicolumn{1}{c}{$\Phi_\mathrm
{SXR}$} &
 \multicolumn{1}{c}{Position} & \multicolumn{1}{c}{$\Delta t_{35}$} & \multicolumn{1}{c}{$F_{35}$} &
 \multicolumn{1}{c}{$\Phi_{35}$} & \multicolumn{1}{c}{$J_{100}$} & \multicolumn{1}{c}{$\Phi_{100}$} &
 \multicolumn{1}{c}{$J_{10}$} & \multicolumn{1}{c}{$\delta_\mathrm{p}$} & \\

&   &   & \multicolumn{1}{c}{class}  &
\multicolumn{1}{c}{$10^{-3}$ J m$^{-2}$} &   &
\multicolumn{1}{c}{min} &
 \multicolumn{1}{c}{$10^{3}$~sfu} & \multicolumn{1}{c}{$10^{5}$~sfu~s} & \multicolumn{1}{c}{pfu} &
 \multicolumn{1}{c}{$10^{3}$~pfu~s} & \multicolumn{1}{c}{pfu} &   & \multicolumn{1}{c}{km s$^{-1}$}  \\

 \multicolumn{1}{c}{(1)} & \multicolumn{1}{c}{(2)} & \multicolumn{1}{c}{(3)} & \multicolumn{1}{c}{(4)} &
 \multicolumn{1}{c}{(5)} & \multicolumn{1}{c}{(6)} & \multicolumn{1}{c}{(7)} & \multicolumn{1}{c}{(8)} &
 \multicolumn{1}{c}{(9)} &  \multicolumn{1}{c}{(10)} & \multicolumn{1}{c}{(11)} & \multicolumn{1}{c}{(12)} &
 \multicolumn{1}{c}{(13)} & \multicolumn{1}{c}{(14)}  \\

 \hline

 \multicolumn{14}{c}{mX events with extreme fluxes at 35 GHz ($F_{35} > 10^4$~sfu)  }\\

1 & 1990-04-15 & 02:55 & X1.4 & 190 & N32E54 & 66 & 20 & 202 & 0.04 & 4 & 9 & 2.35 & U  \\

2 & 1990-05-21 & 22:17 & X5.5 & 120 & N34W37 & 7 & 38 & 45 & 18 & 560 & 300 & 1.22$^1$ & U  \\

3 & 1991-03-22 & 22:45 & X9.4 & 200 & S26E28 & 2 & 122 & 65 & 55 & 1300 & 28000 & 2.70 & U  \\

4 & 1991-03-29 & 06:48 & X2.4 & 70 & S28W60 & 7 & 11 & 10 & 0 & 0 & 20 & U & U  \\

5 & 1991-05-18 & 05:46 & X2.8 & 470 & N32W87 & 26 & 21 & 143 & 0.05 & 3.3 & 7 & 2.14 & U  \\

6 & 1991-06-04 & 03:41 & X12 & 1060 & N30E60 & 15 & 130$^2$ & 470$^2$ & 2 & 90 & 50 & 1.40 & U  \\

7 & 1991-06-06 & 01:12 & X12 & 750 & N33E44 & 17 & 130$^2$ & 890$^2$ & 2.5 & 569 & 200 & 1.90 & U  \\

8 & 1991-06-09 & 01:40 & X10 & 310 & N34E04 & 7 & 74 & 87 & 1.2 & 17 & 80 & 1.82 & U  \\

9 & 1991-06-11 & 02:06 & X12 & 500 & N32W15 & 18 & 46 & 159 & 168 & 2200 & 3000 & 1.25$^1$ & U  \\

10 & 1991-10-24 & 02:41 & X2.1 & 35 & S15E60 & 0.6 & 34 & 6.4 & 0 & 0 & 0 & U & U  \\

11 & 1992-11-02 & 03:08 & X9.0 & 530 & S23W90 & 15 & 41 & 195 & 70 & 2900 & 800 & 1.06$^1$ & U  \\

12 & 2001-04-02 & 21:51 & X17 & 930 & N18W82 & 6 & 25 & 38 & 4.8 & 220 & 380 & 1.90 & 2505  \\

13 & 2002-07-23 & 00:35 & X4.8 & 210 & S13E72 & 17 & 15 & 51 & 0 & 0 & 0 & U & 2285  \\

14 & 2002-08-24 & 01:12 & X3.1 & 178 & S02W81 & 16 & 11 & 46 & 27 & 400 & 220 & 0.91$^1$ & 1913  \\

15 & 2004-11-10 & 02:13 & X2.5 & 80 & N09W49 & 7$^2$ & 10$^2$ & 15$^2$ & 2 & 71 & 75 & 1.57 & 3387  \\

16 & 2005-01-20 & 07:01 & X7.1 & 500 & N12W58 & 25 & 85 & 370 & 680 & 6400 & 1800 & 0.42$^1$ & 2800$^3$  \\

17 & 2006-12-13 & 02:40 & X3.4 & 310 & S06W24 & 31 & 14 & 32 & 88 & 1900 & 695 & 0.89$^1$ & 1774  \\

18 & 2012-03-07 & 00:24 & X5.4 & 310 & N17E15 & 80 & 11 & 136 & 67 & 5300 & 1500 & 1.35 & 2684  \\

19 & 2012-07-06 & 23:08 & X1.1 & 12 & S15W63 & 3 & 17 & 12 & 0.27 & 7.2 & 22 & 1.91 & 1828  \\

20 & 2014-02-25 & 00:49 & X4.9 & 110 & S12E82 & 16 & 48 & 76 & 0.8 & 102.5 & 19 & 1.38 & 2147  \\

 \multicolumn{14}{c}{mS events with strong fluxes at 35 GHz ($10^3$~sfu $ < F_{35} < 10^4$~sfu)  }\\

21 & 1990-05-11 & 05:48 & X2.4 & 70 & N15E13 & 14 & 2.0 & 1.8 & 0 & 0 & 0 & U & U  \\

22 & 1990-05-21 & 01:25 & M4.8 & 20 & N33W30 & 7 & 1.3 & 0.6 & U & U & U & U & U  \\

 \hline
 \end{tabular*}
 \end{table*}

\setcounter{table}{0}
\begin{table*}
 \caption{(\textit{Continued})}
 \begin{tabular*}{\maxfloatwidth}{rcccrcrrrrrrcr}
 \hline

No & Date & \multicolumn{4}{c}{Flare} & \multicolumn{3}{c}{Microwave burst} & \multicolumn{4}{c}{Protons near Earth} &$V_\mathrm{CME}$ \\

& & \multicolumn{1}{c}{$T_\mathrm{peak}$} &
\multicolumn{1}{c}{GOES} & \multicolumn{1}{c}{$\Phi_\mathrm
{SXR}$} &
 \multicolumn{1}{c}{Position} & \multicolumn{1}{c}{$\Delta t_{35}$} & \multicolumn{1}{c}{$F_{35}$} &
 \multicolumn{1}{c}{$\Phi_{35}$} & \multicolumn{1}{c}{$J_{100}$} & \multicolumn{1}{c}{$\Phi_{100}$} &
 \multicolumn{1}{c}{$J_{10}$} & \multicolumn{1}{c}{$\delta_\mathrm{p}$} & \\

&   &   & \multicolumn{1}{c}{class}  &
\multicolumn{1}{c}{$10^{-3}$ J m$^{-2}$} &   &
\multicolumn{1}{c}{min} &
 \multicolumn{1}{c}{$10^{3}$~sfu} & \multicolumn{1}{c}{$10^{5}$~sfu~s} & \multicolumn{1}{c}{pfu} &
 \multicolumn{1}{c}{$10^{3}$~pfu~s} & \multicolumn{1}{c}{pfu} &   & \multicolumn{1}{c}{km s$^{-1}$}  \\

 \multicolumn{1}{c}{(1)} & \multicolumn{1}{c}{(2)} & \multicolumn{1}{c}{(3)} & \multicolumn{1}{c}{(4)} &
 \multicolumn{1}{c}{(5)} & \multicolumn{1}{c}{(6)} & \multicolumn{1}{c}{(7)} & \multicolumn{1}{c}{(8)} &
 \multicolumn{1}{c}{(9)} &  \multicolumn{1}{c}{(10)} & \multicolumn{1}{c}{(11)} & \multicolumn{1}{c}{(12)} &
 \multicolumn{1}{c}{(13)} & \multicolumn{1}{c}{(14)}  \\

 \hline

23 & 1990-05-23 & 04:20 & M8.7 & 50 & N33W55 & 10 & 1.0 & 1.5 & 0 & 0 & 0 & U & U  \\

24 & 1990-06-10 & 07:27 & M2.3 & 7 & N10W10 & 3 & 1.0 & 0.8 & 0 & 0 & 0 & U & U  \\

25 & 1991-01-25 & 06:30 & X10 & 250 & S12E90 & 6 & 9.4 & 13 & 0.14 & 13 & 1 & 0.85 & U  \\

26 & 1991-03-05 & 23:26 & M6.2 & 7 & S23E79 & 2 & 1.4 & 0.6 & 0 & 0 & 0 & U & U  \\

27 & 1991-03-07 & 07:49 & X5.5 & 15 & S20E62 & 3 & 2.0 & 1.5 & 0.08 & 1 & 0.7 & 0.94 & U  \\

28 & 1991-03-13 & 08:04 & X1.3 & 40 & S11E43 & 2 & 3.6 & 0.9 & 0.03 & 0.2 & 4.6 & 2.18 & U  \\

29 & 1991-03-16 & 00:50 & X1.8 & 40 & S10E09 & 3 & 3.2 & 0.4 & 0 & 0 & 0 & U & U  \\

30 & 1991-03-16 & 21:56 & M6.0 & 20 & S09W04 & 4 & 1.6 & 0.2 & 0 & 0 & 0 & U & U  \\

31 & 1991-03-19 & 01:58 & M6.7 & 8 & S10W33 & 1 & 7.2 & 1.9 & 0 & 0 & 0 & U & U  \\

32 & 1991-03-21 & 23:43 & M5.4 & 17 & S25E40 & 3 & 7.2 & 3.4 & 0 & 0 & 0 & U & U  \\

33 & 1991-03-23 & 22:18 & M5.6 & 30 & S25E16 & 15 & 1.7 & 1.1 & U & U & U & U & U  \\

34 & 1991-03-25 & 00:22 & X1.1 & 50 & S26E01 & 11 & 3.9 & 8.8 & 0 & 0 & 0 & U & U  \\

35 & 1991-03-25 & 08:18 & X5.3 & 150 & S25W03 & 4 & 4.2 & 4.1 & 0.5 & 6 & 150 & 2.47 & U  \\

36 & 1991-05-16 & 06:54 & M8.9 & 60 & N30W56 & 9 & 8.0 & 11 & U & U & U & U & U  \\

37 & 1991-05-29 & 23:43 & X1.0 & 20 & N05E38 & 1 & 1.7 & 0.4 & U & U & 0.8 & U & U  \\

38 & 1991-06-30 & 03:01 & M5.0 & 20 & S06W19 & 0.8 & 2.0 & 0.2 & 0 & 0 & 0 & U & U  \\

39 & 1991-07-30 & 07:12 & M7.2 & 6 & N14W58 & 0.9 & 2.0 & 0.3 & 0 & 0 & 0 & U & U  \\

40 & 1991-07-31 & 00:53 & X2.3 & 70 & S17E11 & 5 & 1.6 & 1.6 & 0 & 0 & 0 & U & U  \\

41 & 1991-08-02 & 03:16 & X1.5 & 50 & N25E15 & 8 & 1.2 & 2.7 & 0 & 0 & 0 & U & U  \\

42 & 1991-08-03 & 01:23 & M2.9 & 2.6 & N24E05 & 3 & 2.8 & 0.8 & 0 & 0 & 0 & U & U  \\

43 & 1991-08-25 & 00:49 & X2.1 & 260 & N24E77 & 29 & 1.4 & 10 & 0.03 & 1.1 & 21 & 2.84 & U  \\

44 & 1991-10-27 & 05:49 & X6.1 & 150 & S13E15 & 6 & 8.8 & 13 & 0 & 0 & 40 & U & U  \\

45 & 1991-11-02 & 06:47 & M9.1 & 22 & S13W61 & 3 & 1.4 & 0.5 & 0 & 0 & 0.3 & U & U  \\

46 & 1991-11-15 & 22:38 & X1.5 & 50 & S13W19 & 4 & 1.5 & 1.2 & 0.28 & 2.6 & 1.1 & 0.59 & U  \\

 \hline
 \end{tabular*}
 \end{table*}

\setcounter{table}{0}
\begin{table*}
 \caption{(\textit{Continued})}
 \begin{tabular*}{\maxfloatwidth}{rcccrcrrrrrrcr}
 \hline

No & Date & \multicolumn{4}{c}{Flare} & \multicolumn{3}{c}{Microwave burst} & \multicolumn{4}{c}{Protons near Earth} &$V_\mathrm{CME}$ \\

& & \multicolumn{1}{c}{$T_\mathrm{peak}$} &
\multicolumn{1}{c}{GOES} & \multicolumn{1}{c}{$\Phi_\mathrm
{SXR}$} &
 \multicolumn{1}{c}{Position} & \multicolumn{1}{c}{$\Delta t_{35}$} & \multicolumn{1}{c}{$F_{35}$} &
 \multicolumn{1}{c}{$\Phi_{35}$} & \multicolumn{1}{c}{$J_{100}$} & \multicolumn{1}{c}{$\Phi_{100}$} &
 \multicolumn{1}{c}{$J_{10}$} & \multicolumn{1}{c}{$\delta_\mathrm{p}$} & \\

&   &   & \multicolumn{1}{c}{class}  &
\multicolumn{1}{c}{$10^{-3}$ J m$^{-2}$} &   &
\multicolumn{1}{c}{min} &
 \multicolumn{1}{c}{$10^{3}$~sfu} & \multicolumn{1}{c}{$10^{5}$~sfu~s} & \multicolumn{1}{c}{pfu} &
 \multicolumn{1}{c}{$10^{3}$~pfu~s} & \multicolumn{1}{c}{pfu} &   & \multicolumn{1}{c}{km s$^{-1}$}  \\

 \multicolumn{1}{c}{(1)} & \multicolumn{1}{c}{(2)} & \multicolumn{1}{c}{(3)} & \multicolumn{1}{c}{(4)} &
 \multicolumn{1}{c}{(5)} & \multicolumn{1}{c}{(6)} & \multicolumn{1}{c}{(7)} & \multicolumn{1}{c}{(8)} &
 \multicolumn{1}{c}{(9)} &  \multicolumn{1}{c}{(10)} & \multicolumn{1}{c}{(11)} & \multicolumn{1}{c}{(12)} &
 \multicolumn{1}{c}{(13)} & \multicolumn{1}{c}{(14)}  \\

 \hline

47 & 1992-02-14 & 23:09 & M7.0 & 15 & S12E02 & 1 & 1.0 & 0.3 & 0 & 0 & 0 & U & U  \\

48 & 1992-02-27 & 08:11 & C2.6 & 0.2 & N03W05 & 0.6 & 1.0 & 0.1 & 0 & 0 & 0 & U & U  \\

49 & 1992-06-28 & 05:15 & X1.8 & 170 & N11W90 & 14 & 1.3 & 7.5 & 0.22 & 1.6 & 14 & 1.80 & U  \\

50 & 1994-01-16 & 23:25 & M6.1 & 30 & N07E71 & 9 & 1.2 & 1.1 & 0 & 0 & 0 & U & U  \\

51 & 1997-11-04 & 05:58 & X2.1 & 26 & S14W33 & 3 & 1.0 & 0.7 & 2.3 & 59 & 72 & 1.50 & 785  \\

52 & 1998-08-08 & 03:17 & M3.0 & 2 & N14E72 & 0.7 & 2.0 & 0.5 & 0 & 0 & 0 & U & U  \\

53 & 1998-08-22 & 00:01 & M9.0 & 30 & N42E51 & 6 & 1.0 & 1.7 & U & U & 2.5 & U & U  \\

54 & 1998-11-22 & 06:42 & X3.7 & 100 & S27W82 & 7 & 6.7 & 5.5 & 0.22 & 1.5 & 4 & 1.26 & U  \\

55 & 1999-08-20 & 23:08 & M9.8 & 7 & S23E60 & 1 & 3.0 & 0.6 & 0 & 0 & 0 & U & 812  \\

56 & 1999-12-28 & 00:48 & X4.5 & 100 & N23W47 & 2 & 2.2 & 1.2 & 0.1 & 0.5 & 0.5 & 0.69 & 672  \\

57 & 2000-09-30 & 23:21 & X1.2 & 30 & N09W75 & 4 & 5.3 & 2.3 & 0 & 0 & 0 & U & C  \\

58 & 2000-11-24 & 05:02 & X2.0 & 38 & N19W05 & 2 & 9.3 & 5.6 & 0.58 & 32 & 8 & 1.13 & 1289  \\

59 & 2001-03-10 & 04:05 & X6.7 & 7 & N26W42 & 1 & 1.7 & 0.3 & 0 & 0 & 0.2 & U & 819  \\

60 & 2001-04-03 & 03:57 & X1.2 & 146 & S21E71 & 31 & 2.9 & 19 & 0 & 0 & U & U & 1613  \\

61 & 2001-04-10 & 05:26 & X2.3 & 100 & S24W05 & 30 & 3.0 & 28 & 0.47 & 15 & 100 & 2.32 & 2411  \\

62 & 2001-10-12 & 03:27 & C7.6 & 1 & N16E70 & 1 & 1.3 & 0.3 & 0 & 0 & 0 & U & U  \\

63 & 2001-10-25 & 05:21 & C5.2 & 0.4 & S19W17 & 1 & 1.2 & 0.3 & 0 & 0 & 1 & U & C  \\

64 & 2002-02-20 & 06:12 & M5.1 & 16 & N13W68 & 5 & 1.5 & 0.9 & 0.1 & 0.3 & 10 & 2.00 & 952  \\

65 & 2002-07-18 & 03:37 & M2.2 & 5 & N19W27 & 2 & 1.4 & 0.3 & 0 & 0 & 0 & U & C  \\

66 & 2002-08-20 & 01:40 & M5.0 & 5 & S08W34 & 0.5 & 1.8 & 0.1 & 0  & 0 & 0 & U & 961  \\

67 & 2002-08-21 & 01:41 & M1.4 & 2 & S10W47 & 1 & 1.3 & 0.1 & 0 & 0 & 0 & U & 400  \\

68 & 2002-08-21 & 05:34 & X1.0 & 10 & S09W50 & 0.7 & 1.4 & 0.3 & 0 & 0 & 0 & U & 268  \\

69 & 2003-04-26 & 00:58 & M2.1 & 2.2 & N20W65 & 2 & 2.2 & 0.4 & 0 & 0 & 0 & U & 690  \\

70 & 2003-04-26 & 03:06 & M2.1 & 2.4 & N20W69 & 0.3 & 2.4 & 0.1 & 0 & 0 & 0 & U & 289  \\

 \hline
 \end{tabular*}
 \end{table*}

\setcounter{table}{0}
\begin{table*}
 \caption{(\textit{Continued})}
 \begin{tabular*}{\maxfloatwidth}{rcccrcrrrrrrcr}
 \hline

No & Date & \multicolumn{4}{c}{Flare} & \multicolumn{3}{c}{Microwave burst} & \multicolumn{4}{c}{Protons near Earth} &$V_\mathrm{CME}$ \\

& & \multicolumn{1}{c}{$T_\mathrm{peak}$} &
\multicolumn{1}{c}{GOES} & \multicolumn{1}{c}{$\Phi_\mathrm
{SXR}$} &
 \multicolumn{1}{c}{Position} & \multicolumn{1}{c}{$\Delta t_{35}$} & \multicolumn{1}{c}{$F_{35}$} &
 \multicolumn{1}{c}{$\Phi_{35}$} & \multicolumn{1}{c}{$J_{100}$} & \multicolumn{1}{c}{$\Phi_{100}$} &
 \multicolumn{1}{c}{$J_{10}$} & \multicolumn{1}{c}{$\delta_\mathrm{p}$} & \\

&   &   & \multicolumn{1}{c}{class}  &
\multicolumn{1}{c}{$10^{-3}$ J m$^{-2}$} &   &
\multicolumn{1}{c}{min} &
 \multicolumn{1}{c}{$10^{3}$~sfu} & \multicolumn{1}{c}{$10^{5}$~sfu~s} & \multicolumn{1}{c}{pfu} &
 \multicolumn{1}{c}{$10^{3}$~pfu~s} & \multicolumn{1}{c}{pfu} &   & \multicolumn{1}{c}{km s$^{-1}$}  \\

 \multicolumn{1}{c}{(1)} & \multicolumn{1}{c}{(2)} & \multicolumn{1}{c}{(3)} & \multicolumn{1}{c}{(4)} &
 \multicolumn{1}{c}{(5)} & \multicolumn{1}{c}{(6)} & \multicolumn{1}{c}{(7)} & \multicolumn{1}{c}{(8)} &
 \multicolumn{1}{c}{(9)} &  \multicolumn{1}{c}{(10)} & \multicolumn{1}{c}{(11)} & \multicolumn{1}{c}{(12)} &
 \multicolumn{1}{c}{(13)} & \multicolumn{1}{c}{(14)}  \\

 \hline

71 & 2003-05-28 & 00:27 & X3.6 & 130 & S08W22 & 14 & 3.5 & 10 & 0.15 & 2.6 & 121 & 2.90 & 1366  \\

72 & 2003-05-29 & 01:05 & X1.2 & 40 & S07W31 & 12 & 1.2 & 3.4 & 0.03 & 1.3 & 2 & 1.82 & 1237  \\

73 & 2003-05-31 & 02:29 & M9.3 & 21 & S06W60 & 8 & 3.5 & 13 & 0.8 & 16 & 27 & 1.53 & 1835  \\

74 & 2003-06-15 & 23:56 & X1.3 & 70 & S07E80 & 8 & 1.9 & 5.8 & 0 & 0 & 0 & U & 2053  \\

75 & 2003-06-17 & 22:55 & M6.8 & 40 & S08E58 & 23 & 1.8 & 1.5 & 0 & 0 & 0 & U & 1813  \\

76 & 2003-10-24 & 02:54 & M7.6 & 70 & S19E72 & 32 & 3.9 & 28 & 0 & 0 & 0 & U & 1055  \\

77 & 2003-10-26 & 06:54 & X1.2 & 160 & S17E42 & 62 & 3.6 & 20 & 0 & 0 & 0 & U & 1371  \\

78 & 2004-01-06 & 06:29 & M5.8 & 15 & N05E89 & 8 & 1.0 & 2.3 & 0 & 0 & 0 & U & 1469  \\

79 & 2004-01-07 & 04:04 & M4.5 & 23 & N02E82 & 9 & 1.8 & 4.4 & 0 & 0 & 0 & U & 1581  \\

80 & 2004-07-16 & 02:06 & X1.3 & 25 & S10E39 & 5 & 1.5 & 1.1 & 0 & 0 & 0 & U & 154  \\

81 & 2004-08-14 & 05:44 & M7.4 & 11 & S12W29 & 7 & 1.0 & 0.4 & 0 & 0 & 0 & U & 307  \\

82 & 2004-10-30 & 06:18 & M4.2 & 7 & N13W21 & 7 & 1.3 & 0.7 & 0.04 & 0.3 & 0.9 & 1.35 & 422  \\

83 & 2004-11-03 & 03:35 & M1.6 & 6 & N07E46 & 10 & 1.0 & 3.2 & 0 & 0 & 0.4 & U & 918  \\

84 & 2005-01-01 & 00:31 & X1.7 & 40 & N04E35 & 6 & 1.7 & 3.0 & 0 & 0 & 0 & U & 832  \\

85 & 2005-01-15 & 00:43 & X1.2 & 25 & N13E05 & 6 & 3.3 & 1.5 & 0 & 0 & 0 & U & C  \\

86 & 2005-07-30 & 06:35 & X1.3 & 100 & N11E58 & 27 & 1.2 & 5.1 & 0 & 0 & 0 & U & 1968  \\

87 & 2005-08-25 & 04:44 & M6.4 & 10 & N08E82 & 5 & 4.3 & 5.1 & 0 & 0 & 0 & U & 1327  \\

88 & 2005-09-13 & 23:22 & X1.7 & 33 & S11E10 & 6 & 5.0 & 1.3 & 0.05 & 1 & 200 & 3.6 & 999$^5$  \\

89 & 2005-09-17 & 06:05 & M9.8 & 20 & S11W41 & 6 & 1.3 & 1.8 & 0 & 0 & 1.4 & U & C  \\

90 & 2010-06-12 & 00:57 & M2.0 & 4 & N24W47 & 2$^2$ & 4.0$^2$ & 2.7$^2$ & 0.05 & 0.26 & 0.9 & 1.26 & 486  \\

91 & 2011-08-04 & 03:57 & M9.3 & 26 & N16W49 & 11 & 1.4 & 3.4 & 1.5 & 28 & 77 & 1.71 & 1315  \\

92 & 2011-08-09 & 08:05 & X6.9 & 86 & N17W83 & 6 & 1.0 & 4.4 & 2.5 & 22 & 22 & 0.94 & 1610  \\

93$^4$ & 2011-09-06 & 22:20 & X2.2 & 50 & N14W18 & 2 & 3.0 & 5.5 & 0.5 & 8.2 & 10 & 1.30 & 575  \\

94$^4$ & 2011-09-07 & 22:37 & X1.7 & 44 & N14W28 & 2 & 1.0 & 4.6 & 0.05 & 1.3 & 10 & 2.30 & 792  \\

 \hline
 \end{tabular*}
 \end{table*}

\setcounter{table}{0}
\begin{table*}
 \caption{(\textit{Continued})}
 \begin{tabular*}{\maxfloatwidth}{rcccrcrrrrrrcr}
 \hline

No & Date & \multicolumn{4}{c}{Flare} & \multicolumn{3}{c}{Microwave burst} & \multicolumn{4}{c}{Protons near Earth} &$V_\mathrm{CME}$ \\

& & \multicolumn{1}{c}{$T_\mathrm{peak}$} &
\multicolumn{1}{c}{GOES} & \multicolumn{1}{c}{$\Phi_\mathrm
{SXR}$} &
 \multicolumn{1}{c}{Position} & \multicolumn{1}{c}{$\Delta t_{35}$} & \multicolumn{1}{c}{$F_{35}$} &
 \multicolumn{1}{c}{$\Phi_{35}$} & \multicolumn{1}{c}{$J_{100}$} & \multicolumn{1}{c}{$\Phi_{100}$} &
 \multicolumn{1}{c}{$J_{10}$} & \multicolumn{1}{c}{$\delta_\mathrm{p}$} & \\

&   &   & \multicolumn{1}{c}{class}  &
\multicolumn{1}{c}{$10^{-3}$ J m$^{-2}$} &   &
\multicolumn{1}{c}{min} &
 \multicolumn{1}{c}{$10^{3}$~sfu} & \multicolumn{1}{c}{$10^{5}$~sfu~s} & \multicolumn{1}{c}{pfu} &
 \multicolumn{1}{c}{$10^{3}$~pfu~s} & \multicolumn{1}{c}{pfu} &   & \multicolumn{1}{c}{km s$^{-1}$}  \\

 \multicolumn{1}{c}{(1)} & \multicolumn{1}{c}{(2)} & \multicolumn{1}{c}{(3)} & \multicolumn{1}{c}{(4)} &
 \multicolumn{1}{c}{(5)} & \multicolumn{1}{c}{(6)} & \multicolumn{1}{c}{(7)} & \multicolumn{1}{c}{(8)} &
 \multicolumn{1}{c}{(9)} &  \multicolumn{1}{c}{(10)} & \multicolumn{1}{c}{(11)} & \multicolumn{1}{c}{(12)} &
 \multicolumn{1}{c}{(13)} & \multicolumn{1}{c}{(14)}  \\

 \hline

95 & 2012-01-23 & 03:59 & M8.7 & 60 & N29W36 & 39 & 2.2 & 24 & 2.3 & 85 & 2700 & 3.07 & 2175  \\

96 & 2012-10-23 & 03:17 & X1.8 & 40 & S13E58 & 3 & 4.4 & 1.5 & 0 & 0 & 0 & U & C  \\

97 & 2013-05-13 & 02:17 & X1.7 & 130 & N10E89 & 18 & 1.2 & 5.2 & 0 & 0 & 0 & U & 1270  \\

98 & 2013-05-14 & 01:11 & X3.2 & 100 & N11E74 & 22 & 1.1 & 0.6 & 0.03 & 0.12 & 1 & 1.52 & 2625  \\

99 & 2013-10-28 & 01:59 & X1.0 & 61 & N04W66 & 8 & 1.9 & 2.6 & 0.12 & 7 & 5 & 1.62 & 695  \\

100 & 2013-11-08 & 04:26 & X1.1 & 20 & S44E86 & 4 & 4.0 & 3.0 & 0 & 0 & 0.7 & U & 497  \\

101 & 2014-10-22 & 01:59 & M8.7 & 80 & S12E21 & 8 & 1.6 & 7.8 & 0 & 0 & 0 & U & C  \\

102 & 2014-10-30 & 00:37 & M1.3 & 20 & S14W81 & 1.1 & 1.4 & 0.2 & 0 & 0 & 0 & U & C  \\

103 & 2014-12-20 & 00:28 & X1.8 & 83 & S18W28 & 6 & 1.0 & 5.2 & 0 & 0 & 1.5 & U & U  \\

104 & 2015-03-10 & 03:24 & M5.1 & 12 & S16E34 & 4.2 & 2.0 & 0.5 & 0 & 0 & 0 & U & U  \\

 \multicolumn{14}{c}{mM events with strong proton fluxes ($J_{100} > 10$ pfu, $10^2$~sfu $< F_{35} < 10^3$~sfu) }\\

105 & 2000-11-08 & 23:28 & M7.8 & 66 & N10W75 & 53 & 0.09 & 2.1 & 320 & 13000 & 14000 & 1.64 & 1738  \\

106 & 2001-12-26 & 05:40 & M7.1 & 110 & N08W54 & 26 & 0.78 & 8.2 & 47 & 600 & 700 & 1.17$^1$ & 1446  \\

107 & 2002-04-21 & 01:51 & X1.5 & 280 & S14W84 & 120 & 0.5$^2$ & 7.2$^2$ & 20 & 1500 & 2000 & 2.0 & 2393  \\

108 & 2012-05-17 & 01:47 & M5.1 & 31 & N09W74 & 17 & 0.2 & 1.7 & 18 & 305 & 230 & 1.11$^1$ & 1582  \\

 \multicolumn{14}{c}{mO backside events with strong proton fluxes ($J_{100} > 10$ pfu)}\\

109 & 1990-05-28 & 04:33 & U & 0 & N36W120 & U & U & U & 4.5 & 295 & 44 & 0.99$^1$ & U  \\

110 & 2001-04-18 & 02:14 & C2.2 & 45 & S20W115 & U & U & U & 12 & 270 & 230 & 1.28$^1$ & 2465  \\

111 & 2001-08-15 & 23:50 & U & 0 & N01W120 & U & U & U & 27 & 670 & 470 & 1.24 & 1575  \\

 \hline
 \end{tabular*}

$^1$GLE event

$^2$Estimated from different data

$^3$A compromise between the estimates of
\inlinecite{Gopalswamy2005} and \inlinecite{Grechnev2008}

$^4$Missing in the NoRP event list

$^5$Excluded from the analysis because of overlap with a preceding
event

 \end{table*}

 \end{kaprotate}

Column (1) of Table~\ref{T-data_table} presents the event number.
Columns (2) and (3) show the date and time of the flare peak
according to GOES reports. Columns (4)--(6) contain GOES class,
start-to-peak SXR fluence, and flare coordinate.

Columns (7)--(9) list the half-height duration, peak intensity, and
total microwave fluence at 35 GHz, $\Phi_{35}$. NoRP records at
35~GHz were absent or damaged for some events. In such cases, the
value of $F_{35}$ was estimated by means of interpolation from the
adjacent frequencies of 17 and 80 GHz and/or from the 34~GHz data of
the Nobeyama Radioheliograph (NoRH; \opencite{Nakajima1994}).

Columns (10)--(13) list parameters of near-Earth proton
enhancements: the peak flux of protons with energies above 100~MeV,
$J_{100}$; the total fluence, $\Phi_{100}$; the peak flux of protons
with energies above 10~MeV, $J_{10}$; the index of the integral
energy proton spectrum, $\delta_\mathrm{p} =
\log_{10}(J_{10}/J_{100})$, which was calculated from the peak
fluxes of protons with different energies occurring at different
times, thus attempting to take account of their velocity dispersion.
The events marked in column (13) with a superscript $(^1)$ were
associated with GLEs. Column~(14) presents the CME speed. Unknown or
uncertain parameters are denoted by `U'. Confined flares are denoted
by `C'.

\begin{figure} 
  \centerline{\includegraphics[width=\textwidth]
   {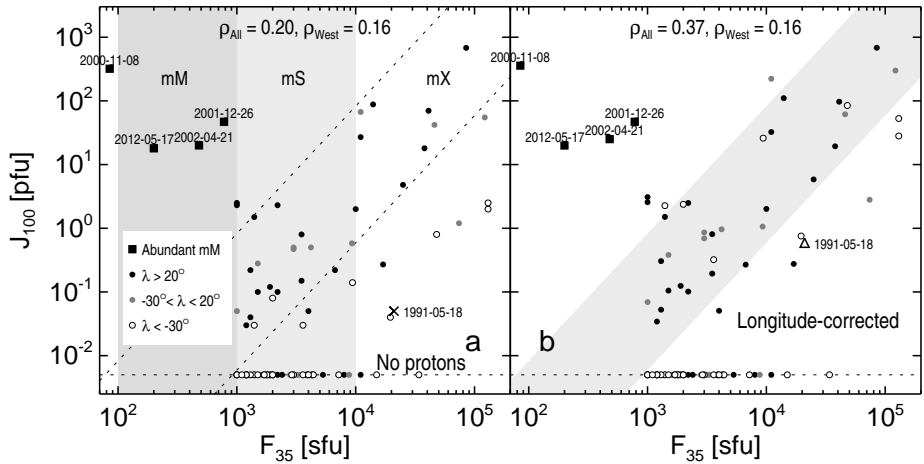}
  }
  \caption{
Peak fluxes of $> 100$~MeV protons versus peak microwave fluxes at
35 GHz: (a)~actual values, (b)~with a positional correction. The
longitude of the solar source for each data point is coded by the
symbols explained in the legend in panel (a). The filled squares
denote the mM events with atypically high SPEs. The 64 events
without detectable proton fluxes falling outside the plot region are
schematically presented at the horizontal dotted line below. The
Pearson correlation coefficients specified in each panel were
calculated separately for all 44 proton events ($\rho_\mathrm{All}$)
and for 26 West proton events only ($\rho_\mathrm{West}$). The
slanted dotted lines, $(F_{35}/13000)^2$ and $(F_{35}/1100)^2$, in
panel (a) and the corresponding shading in panel (b) enclose the
majority of data points (`main sequence'), indicating a direct
relation between the observed $F_{35}$ and $J_{100}$. (Adapted from
Grechnev \textit{et al.}, 2013b).}
  \label{F-long-cor}
  \end{figure}

The data from Table~\ref{T-data_table} are shown in
Figure~\ref{F-long-cor}a, similar to a corresponding figure in
\inlinecite{Grechnev2013b}. For clarity, solar events are
categorized according to their heliolongitude, $\lambda$, into
three intervals with boundaries of $-30^{\circ}$ and
$+20^{\circ}$, presented by the colored circles. The events
without detectable proton fluxes are shown at the horizontal
dotted line below, to reveal their amount. The majority of SPEs is
grouped between the slanted lines $(F_{35}/1100)^{2}$ and
$(F_{35}/13000)^{2}$~pfu, forming the `main sequence'. Four
atypical proton-abundant mM events denoted by the black squares
reside in the upper-left part of the figure, much higher than the
`main sequence'. The correlation coefficients between the
logarithms of the peak values of the microwave and proton fluxes
for all events, $\rho_\mathrm{All}$, and separately for west
events only, with a heliolongitude $\lambda>20^{\circ},
\rho_\mathrm{West}$, are shown in the upper part of the figure.
The correlation for the west events is lower due to a considerable
contribution from the four abundant events, all of which had west
locations, while only 60\% of all SPEs had sources with $\lambda
> 20^{\circ}$.

Following \inlinecite{Kahler1982}, we show in
Figure~\ref{F-long-cor}b the same events, but with a correction
$\exp{\{[(\lambda - 54^\circ)/63]^2\}}$ for the longitudinal
dependence of $> 100$~MeV protons \cite{Belov2009}. This dependence
is close to the result of \inlinecite{Lario2013} obtained for
protons in an energy range of 25--53~MeV. The correction was
formally applied to all events, including west events. The strongest
effect of the correction is for far-east events (open circles), and
somewhat weaker effect is for moderately east events (gray filled
circles). The usage of the longitudinal correction increases the
correlation for the whole ensemble of events by 70\%. We therefore
apply this correction in further analysis to all parameters of
proton enhancements, even if their solar sources had west locations.
Since this correction is uncertain, we additionally considered the
correlation coefficients for west events only.

It is not obvious how to handle the atypical event 5. It is an
outlier with its actual longitude (the slanted cross in
Figure~\ref{F-long-cor}a). According to its properties, event 5
should be handled in a way similar to the east events, but a
suitable correction is unknown. As a tolerable, but practically
inappropriate compromise, we handle this event, as if it had a
middle east longitude of E45. The corresponding triangle in
Figure~\ref{F-long-cor}b shows that this correction is not
excessive.

All of the calculated correlation coefficients and regression
parameters refer to logarithms of analyzed quantities rather than
actual values because of their wide ranges. This way of
linearization allows one to use linear correlation analysis widely
applied in many studies mentioned. On the other hand, logarithms of
zero values are infinite that requires a separate analysis of such
terms. In addition, applying linear statistical methods to
logarithms inevitably results in biased estimates due to the strong
nonlinearity. Thus, the usage of the logarithmic scale is a
necessary compromise, which allows comparing and quantifying
statistical trends of an analyzed quantity on various parameters,
but it is not mathematically rigorous.

\subsection{Peak Flux at 35 GHz and the Probability of a High-Energy SPE}
 \label{S-peak_flux}

The percentage of SPEs associated with mX bursts is 90\%. Protons
$>100$~MeV were observed in 85\% of the mX events. GLEs occurred
after 30\% of the mX events; besides the mX events, GLEs only
occurred after two abundant mM events and two far-side mO events.
The probability of a proton enhancement after an mS burst is
considerably lower, 52\% for $E_{p} > 10$~MeV and 35\% for $E_{p}>
100$ MeV. None of the mS events produced a GLE.

It is difficult to evaluate the probability of SPEs after mM bursts
because of their large number (270) and insufficient accuracy of the
software, which calculates the parameters of the bursts posted at
the NoRP Web site. An accurate processing about 600 events is needed
for a correct evaluation of the probability. Instead of this, we
roughly estimate the upper and lower boundaries for the probability,
using these lists and a catalog of SPEs presented at
\url{http://umbra.nascom.nasa.gov/SEP/}. The total number of proton
events with $J_{10}> 10$~pfu in the catalog from 1990 to March 2015,
whose solar sources fell into the observing time in Nobeyama or were
uncertain, was 70. Protons $> 100$~MeV were not observed in all of
these events. Proceeding from the number of events in the catalogs,
the probability of $> 10$~MeV SPEs after mM bursts was estimated to
be within (8--23)\%. The probability of high-energy SPEs after mM
bursts was somewhat lower.

\begin{figure} 
  \centerline{\includegraphics[width=0.5\textwidth]
   {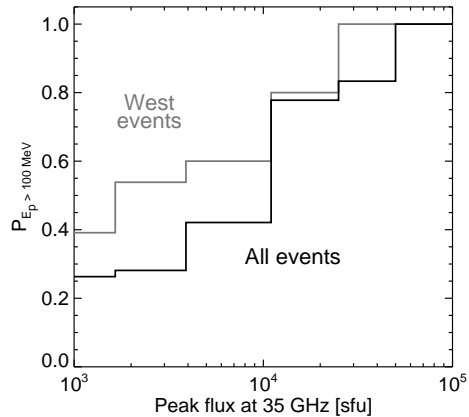}
  }
  \caption{Probability of a near-Earth proton enhancement
with $E_\mathrm{P} > 100$~MeV \textit{vs.} peak flux of the 35 GHz
burst irrespective of the burst duration or the position of a
solar source.}
  \label{F-mw_flux_dist}
  \end{figure}

Figure~\ref{F-mw_flux_dist} presents a more detailed probability
distribution of high-energy SPEs depending on the 35 GHz peak
flux, $F_{35}$. The shape of the histogram is sensitive to the
bins because of a relatively small number of events. The intervals
were chosen to reach a possibly larger number of bins, keeping the
histogram monotonic. After a microwave burst with a peak flux of
$F_{35} \approx 10^{3}$~sfu, the SPE probability is 25--40\%. With
an increase of $F_{35}$, the probability increases, approaching
100\% for $F_{35} > 5 \times 10^{4}$~sfu. The SPE probability
after a west solar event is 10--20\% higher than the probability
averaged over the whole set of events.

Thus, the probability of a proton enhancement directly depends on
the peak flux of the microwave burst at 35 GHz. This fact is
consistent with a result of \inlinecite{Dierckxsens2015} and the
conclusion of \inlinecite{Grechnev2013b} that a powerful microwave
burst indicates at a large proton event with a hard spectrum, up
to a GLE, if the duration of the burst is long. The last condition
is analyzed in the next section.

\subsection{The Role of Duration of Microwave Burst}
 \label{S-durations}

The distribution of SPE peak fluxes \textit{vs.} the peak fluxes
of the microwave bursts and their durations obtained by
\inlinecite{Grechnev2013b} confirmed the well-known reduced proton
productivity of short-duration events. However, this distribution
does not resemble two separate clusters with different durations
that could be expected as a manifestation of two different
acceleration mechanisms. Instead, a general, although scattered,
tendency is surmised. To find the reasons for the reduced proton
productivity of impulsive events, we firstly consider the
properties of the distribution of microwave bursts on their
duration, and in the next section we analyze the correlations
between various combinations of the peak values and fluences of
microwave bursts and proton enhancements.

The events without detectable proton fluxes are presented at the
horizontal dotted line at the bottom of Figure~\ref{F-long-cor}. The
half-height durations of the corresponding microwave bursts range
from 0.3 to 62~min with an average of 7.7~min ($\sigma_\mathrm{n} =
10.2$~min). The durations of SPE-related events range from 2 to
80~min with an average of 14.1 min ($\sigma_\mathrm{p} = 16.5$~min).
The difference in the durations by a factor of 1.8 appears only in
averages, and its significance is questionable.

Let us consider the properties of the distributions of microwave
bursts with $F_{35} \geq 10^3$~sfu on their durations for
SPE-related and non-SPE-related events separately irrespective of
their other parameters. The histograms of these distributions
calculated in a straightforward way are inconclusive because of
the relatively small number of events. We have used for the
analysis a different way of calculating the integral probability
distribution, $P(\Delta t \le t)$. It is an antiderivative of the
histogram with a maximum normalized to unity and characterizes the
probability of an event, if its duration $\Delta t$ does not
exceed a value of $t$.

\begin{figure} 
  \centerline{\includegraphics[height=0.75\textheight]
   {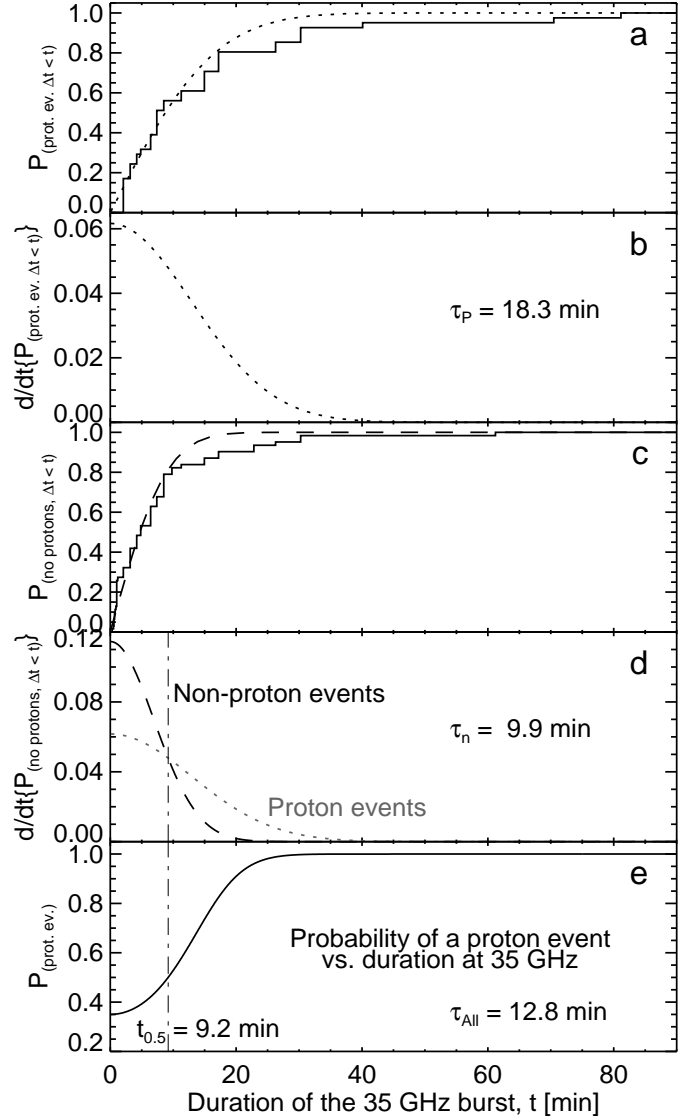}
  }
  \caption{Probability distributions of microwave bursts with peak fluxes
$F_{35} \geq 10^3$~sfu on their durations. (a)~Integral and
(b)~differential probability distributions for SPE-related events;
(c) and (d)~same for non-SPE-related events; (e)~probability of a
proton enhancement depending on the duration of the microwave burst.
The histogram-style lines represent actual distributions, and the
dotted curve show the analytic fit. The gray dotted curve in panel
(d) corresponds to the analytic fit in panel (b). The vertical
dash-dotted line in panels (d) and (e) corresponds to the 50\%
probability ($t_{35} = 9.2$~minutes).}
  \label{F-dur_dist}
  \end{figure}

The solid histogram-like line in Figure~\ref{F-dur_dist}a presents
the integral probability distribution for SPE-related microwave
bursts depending on their duration. This distribution in the linear
and logarithmic representations seems to be close to the error
function, $\mathrm{erf}(t/\tau)$, indicating that the duration
distribution is close to normal. The derivative of the integral
distribution $P(\Delta t\le t) = \mathrm{erf}(t/\tau)$ is a
probability density function, which is known to be a Gaussian
centered at zero, $2\exp\{-(t/\tau)^2\}/(\tau\sqrt{\pi})$. By
minimizing the difference between the actual distribution and the
fit, we have found $\tau_\mathrm{p} = 18.3$~min, which characterizes
a typical duration of a SPE-related microwave burst. The
corresponding fitting functions are shown by the dotted lines in
Figures \ref{F-dur_dist}a and \ref{F-dur_dist}b.

The character of the duration distribution for non-SPE-related
events turned out to be the same (Gaussian centered at zero), but
with a lesser width, $\tau_\mathrm{n} = 9.9$~min (Figures
\ref{F-dur_dist}c and \ref{F-dur_dist}d). The fit is shown by the
dashed line. For comparison, the gray dotted line in
Figure~\ref{F-dur_dist}d is the fit of the distribution for proton
events. The ratio of the widths of the distributions for proton and
non-proton events, $\tau_\mathrm{p} / \tau_\mathrm{n} = 1.8$,
corresponds to the ratio of their actual average durations.

A calculated ratio of the probabilities of proton and non-proton
events in Figure~\ref{F-dur_dist}d is $P_\mathrm{p}/P_\mathrm{n} =
\tau_\mathrm{p}/\tau_\mathrm{n}
\,\exp\{{t_{35}}^2\,(1/\tau_\mathrm{p}^2 -
1/\tau_\mathrm{n}^2)\}$. If the duration of a microwave burst,
$t_{35}$, is known, then the probability of a proton enhancement,
$P_\mathrm{p}(t_{35})$, can be estimated as
 \begin{equation}
P_\mathrm{p}(t_{35}) = 1/(P_\mathrm{p}/P_\mathrm{n}+1) =
(\tau_\mathrm{p}/\tau_\mathrm{n}\,\exp\{t_{35}^{2}(1/\tau_\mathrm{p}^{2}
- 1/\tau_\mathrm{n}^{2})\}+1)^{-1}.
 \label{E-mw_dur_prob}
 \end{equation}
With the parameters found for the analyzed set of events, this
equation gives an estimate of 52\%. Actually, SPEs occurred after 41
out of 103 microwave bursts with fluxes $F_{35} \geq 10^{3}$~sfu,
\textit{i.e.}, in $\approx 40\%$ of the events. The calculated
probability for proton events \textit{vs.} the duration of the
35~GHz burst is shown in Figure~\ref{F-dur_dist}e. The vertical
dash-dotted line in Figures \ref{F-dur_dist}d and \ref{F-dur_dist}e
denotes the burst duration of $t_{0.5} = 9.2$~min, at which the
distribution functions of the proton and non-proton events are
equal, that corresponds to the probability of 0.5. According to
Figure~\ref{F-dur_dist}e, if the peak flux of the 35 GHz burst
exceeds $10^{3}$~sfu and its duration exceeds 30~min, then the SPE
probability is close to 100\%.

The identity of the distribution functions for SPE-related and
non-SPE-related events indicate the absence of essential differences
between these classes of the events manifesting in their durations.
The duration distribution of the whole set of microwave bursts,
including both proton and non-proton events, is also close to the
normal distribution with $\tau_\mathrm{All} = 12.8$~min. This
distribution is not associated in any way with the proton
productivity of the events, being an intrinsic characteristic of
microwave bursts. A probable reason for the different widths of the
distributions, $\tau_\mathrm{n} < \tau_\mathrm{All} <
\tau_\mathrm{p}$, is the sensitivity of the detectors, which measure
the proton fluxes in the Earth orbit against the radiation
background. The decrease of the SPE peak due to the velocity
dispersion of the proton bunch in the interplanetary space and other
propagation effects is particularly strong, if the bunch has a short
duration.

The velocity dispersion (SPE rise) time for $> 100$~MeV protons can
be roughly estimated as a difference between the straight Sun--Earth
propagation times of the 100~MeV and relativistic protons,
 $t_\mathrm{D} \approx 1\, \mathrm{AU} \times (1/v - 1/c) = 1\,
\mathrm{AU}/c \times (1/\sqrt{1-1/(E/m_\mathrm{p}+1)^2} - 1)
\approx 11\ \mathrm{min},$
 with $E = 100$~MeV, $v$, and $m_\mathrm{p}$ being the kinetic
energy, velocity, and the rest mass of protons; $c$ is the speed
of light. A high-energy rollover of the proton spectrum decreases
$t_\mathrm{D}$, while the actual path length, including the Parker
spiral and particularities of the propagation in the
interplanetary space, increases $t_\mathrm{D}$. Thus,
$t_\mathrm{D} \sim t_\mathrm{n}$, consistent with our assumption.
Similar reasons might also control the dependence of the SPE
probability on the peak of the microwave burst shown in
Figure~\ref{F-mw_flux_dist}.

The absence of two different clusters in the durations of the
events, the absence of characteristic durations of the 35~GHz
bursts in the events with protons and without them, and, instead,
the same shapes of their distributions with most probable zero
durations are not consistent with the two distinct classes of
`impulsive' and `gradual' events. Therefore, a possible reason for
the dependence of the number of high-energy protons on the
duration of the event can be not a difference in the particle
acceleration mechanisms, but the duration of the acceleration
process, on which the proton fluence should be directly dependent
in any case. It seems also reasonable to consider the microwave
fluence in addition to the peak flux. The fluence is an energy
characteristic of the microwave emission throughout the flare,
while the peak flux characterizes the maximum of its power
spectral density observed in the event. The correlations between
various combinations of peak values and fluences of the microwave
bursts and proton enhancements are analyzed in the next section.

\subsection{Microwave and Proton Fluences}
 \label{S-fluences}

\begin{figure} 
  \centerline{\includegraphics[width=\textwidth]
   {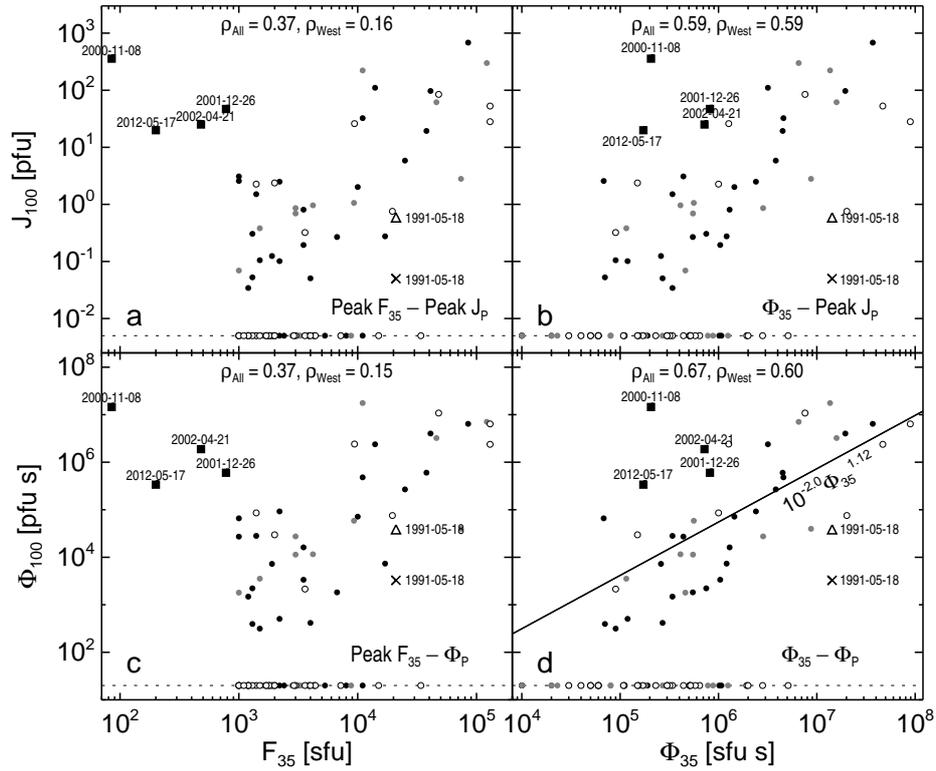}
  }
  \caption{Statistical relations between different combinations
of the peak fluxes and fluences of microwave bursts, on the one
hand, and those of longitude-corrected SPEs, on the other hand.
The Pearson correlation coefficients specified in each panel were
calculated separately for all 44 SPEs presented
($\rho_\mathrm{All}$) and for 26 West events only
($\rho_\mathrm{West}$). The meaning of the symbols is the same as
in Figure~\ref{F-long-cor}. The line in panel (d) represents the
linear fit of the (log--log) distribution. The poorly connected 18
May 1991 event is shown both with a correction (triangle) and
without it (slanted cross).}
  \label{F-plot4}
  \end{figure}

The relations between various combinations of the peak fluxes and
fluences of microwave bursts and proton enhancements are presented
in Figure~\ref{F-plot4}. The correlation coefficients for all events
and, separately, for west events only are shown in the upper parts
of the plots. The poorly connected event 5 was treated with a
correction described in Section~\ref{S-data}. This increased the
correlation coefficients only insignificantly (\textit{e.g.},
$\rho_\mathrm{All}$ from 0.64 to 0.67 in Figure~\ref{F-plot4}d).

The scatter of the data points in the top and bottom left panels is
similar. The four abundant events deviate from the main cloud of
points considerably less in the right panels, where the argument is
$\Phi_{35}$, than in the left panels, where the argument is
$F_{35}$. The main cloud of points without the four abundant events
is narrower in Figure~\ref{F-plot4}d than in Figure~\ref{F-plot4}b.
The best correspondence between the proton and microwave fluences is
confirmed by the highest correlation coefficient of 0.67 for this
combination of the parameters. Note that \inlinecite{Kahler1982}
found the microwave fluences to correlate with peak proton fluxes
higher than the BFS hypothesis predicted (which corresponds to our
Figure~\ref{F-plot4}b), but he did not consider the relation between
the microwave and proton fluences (Figure~\ref{F-plot4}d).

Extraordinarily high fluxes of high-energy protons were observed
in four abundant mM events selected by our criteria. It is
possible that these events were essentially different from the
others. As Figure~\ref{F-plot4} shows, the highest correlation
between the microwave and proton fluences can be due to the long
duration of the abundant events. However, even for the whole set
of events with $F_{35} > 1000$~sfu without the abundant events,
the correlation coefficient between the fluences is 0.82, and does
not exceed 0.75 for other combinations. Thus, the correlation
between the microwave and proton fluences is highest in any case.

The linear fit for the whole data set is $\Phi_{p} = 10^{-1.99 \pm
1.18} \Phi_{35}^{1.12 \pm 0.19}$, and the correlation coefficients
are $\rho_\mathrm{All} = 0.67$, $\rho_\mathrm{West} = 0.60$. The
data set without the abundant mM events and atypical event 5 is
fit with $\Phi_{p} = 10^{-4.17 \pm 0.96} \Phi_{35}^{1.44 \pm
0.15}$, and the correlation coefficients are $\rho_\mathrm{All} =
0.84$, $\rho_\mathrm{West} = 0.91$. The nonlinearity of the
relation might be due to the complex dependence of the
gyrosynchrotron emission on the parameters of radiating electrons,
including their spectral and spatial distributions, magnetic field
strength, and other factors \cite{DulkMarsh1982, Kundu2009}. Our
choice of a high frequency of 35~GHz simplifies the situation; the
scatter at a lower frequency can be wider due to the influence of
these factors.

Because the probability of a detectable $> 100$~MeV SPE depends
directly on the peak flux and duration of a 35 GHz burst
(Figure~\ref{F-mw_flux_dist} and Equation~\ref{E-mw_dur_prob}),
its relation with the 35~GHz fluence in
Figure~\ref{E-mw_fluence-prob} is pronounced still clearer in the
histograms with nearly equal bins covering a range of three orders
of magnitude. The distributions can be approximately fitted by the
empirical relations
\begin{eqnarray}
P_\mathrm{All}(\mathrm{SPE},\ E_\mathrm{p} > 100\ \mathrm{MeV})
\approx 1-\exp\{-[\Phi_{35}/(1.5 \times 10^6)]^{0.5}\}\\ \nonumber
P_\mathrm{West}(\mathrm{SPE},\ E_\mathrm{p} > 100\ \mathrm{MeV})
\approx 1-\exp\{-[\Phi_{35}/(2.7 \times 10^5)]^{0.75}\}.
\label{E-mw_fluence-prob}
\end{eqnarray}
In future, when calibrated microwave measurements would be
available in real time, such relations could be used to promptly
forecast the probability and importance of nearing SPEs with
ongoing update of the quantities issued.

  \begin{figure} 
  \centerline{\includegraphics[width=0.5\textwidth]
   {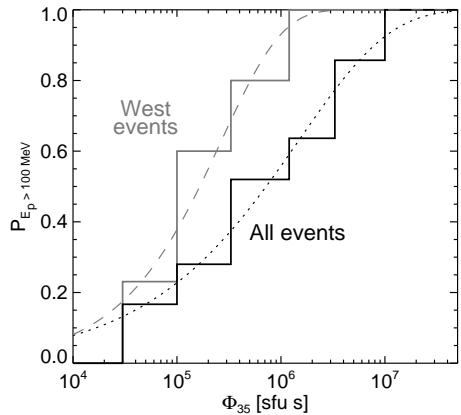}
  }
  \caption{Probability of a near-Earth proton enhancement
with $E_\mathrm{P} > 100$~MeV \textit{vs.} the microwave fluence
at 35 GHz irrespective of any other parameters.}
  \label{F-mw_fluence_dist}
  \end{figure}

\section{Origin of High-Energy SPEs}
 \label{S-origin}

The direct relation between the microwave and proton fluences
indicates the dependence of the total number of high-energy
protons arriving at the Earth orbit on the total duration of the
acceleration process. The correspondence between the durations of
the acceleration process and the microwave burst is obvious, but
it is more difficult to expect such a correspondence, if the
protons are accelerated by shock waves far away from a flare
region. Thus, the results of the preceding
Section~\ref{S-fluences} favor predominantly flare-related origin
of the analyzed SPEs ($F_{35} \geq 10^3$~sfu) with respect to the
hypothesis of their exceptional shock-acceleration. A contribution
from shock-acceleration is also possible, but with a lesser
statistical significance --- probably, in the abundant events. The
suggestion is consistent with the preliminary conclusion of
\inlinecite{Trottet2015} deduced from a different approach. The
authors analyzed correlations between peak proton fluxes and
parameters of flares and CMEs. To additionally verify our
statistical conclusions, we will apply their approach to our data
set.

\subsection{Relations Between Parameters of Eruptive Solar Activity
and Proton Fluences}
 \label{S-relations}

\inlinecite{Trottet2015} analyzed 44 SPEs in an energy range of
15--40 MeV (and corresponding fluxes of subrelativistic electrons)
associated with flares of M and X GOES classes that occurred in
1997--2006 in the west solar hemisphere. The authors calculated
correlation coefficients between logarithms of peak proton fluxes
and parameters characterizing the flares and CMEs. The analyzed
parameters were the peak flux of the SXR emission, start-to-peak
SXR fluence, microwave fluence, and CME speed.

In the proton energy range of 15--40 MeV analyzed by
\inlinecite{Trottet2015}, it is difficult to filter out the
contribution from the acceleration by interplanetary shock waves
far away from the Sun that is, most likely, considerably less for
proton energies above 100~MeV. As Figure~\ref{F-plot4} and the
related text show, the microwave fluence, $\Phi_\mathrm{35}$,
correlates with total proton fluence, $\Phi_\mathrm{SXR}$,
considerably better than with the peak proton flux, $J_{p}$.
Therefore, we analyze the correlations with total fluences of SPEs
rather than their peak fluxes.

Systematic information about CMEs and their plane-of-the-sky speeds
is available in the CME catalog for events since 1996
(\opencite{Yashiro2004}; \url{http://cdaw.gsfc.nasa.gov/CME_list/}).
The speeds listed in the CME catalog are measured for the fastest
feature, and therefore $V_\mathrm{CME}$ for fast CMEs are most
likely related to shock waves (see, \textit{e.g.},
\opencite{Ciaravella2006}). The halo shock fronts ahead of expanding
fast CMEs should have the shapes close to spheroidal or elliptical
ones (\citeauthor{Grechnev2011}, \citeyear{Grechnev2011,
Grechnev2013a, Grechnev2014}; \opencite{Kwon2014};
\opencite{Kwon2015}), and therefore the plane-of-the-sky speeds
measured in the catalog should not be drastically different from the
modules of their vectors. The CME speeds are known for 28 proton
events listed in Table~\ref{T-data_table}.

\begin{figure} 
  \centerline{\includegraphics[width=\textwidth]
   {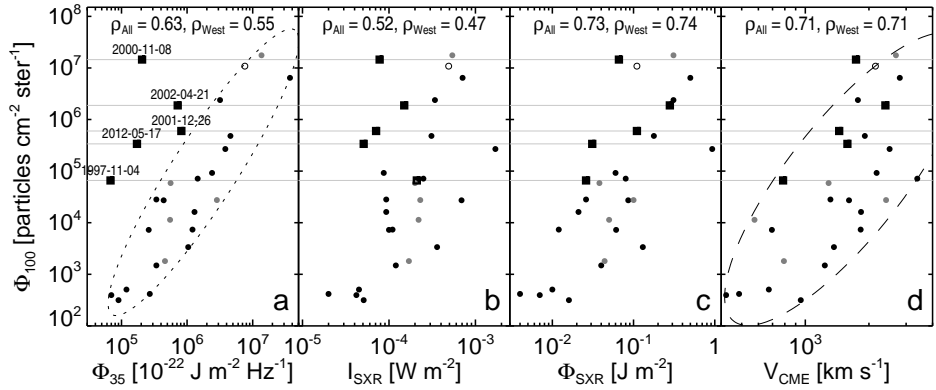}
  }
  \caption{Scatter (log--log) plots of longitude-corrected SPE fluence,
$\Phi_{100}$, versus microwave fluence. $\Phi_{35}$, peak SXR
flux, $I_\mathrm{SXR}$, start-to-peak SXR fluence,
$\Phi_\mathrm{SXR}$, and CME speed, $V_\mathrm{CME}$. The Pearson
correlation coefficients specified in each panel were calculated
separately for all 28 events presented ($\rho_\mathrm{All}$) and
for 22 West events only ($\rho_\mathrm{West}$). The meaning of the
symbols is the same as in Figure~\ref{F-long-cor}. The gray
horizontal lines trace the fluences in the abundant events. The
broken ellipses in panels (a) and (d) enclose all but abundant
events.}
  \label{F-cor4}
  \end{figure}

Figure~\ref{F-cor4} shows the logarithmic scatter plots of the
proton fluence above 100~MeV with the longitudinal correction,
$\Phi_\mathrm{100}$, \textit{vs.} total microwave fluence,
$\Phi_\mathrm{35}$ (Figure~\ref{F-cor4}a); SXR peak flux,
$I_\mathrm{SXR}$ (Figure~\ref{F-cor4}b); its start-to-peak SXR
fluence, $\Phi_\mathrm{SXR}$ (Figure~\ref{F-cor4}c); and the CME
speed, $V_\mathrm{CME}$ (Figure~\ref{F-cor4}d). The events without
SPEs, whose logarithms are infinite, are not included in the
correlation analysis. These events presented in
Figure~\ref{F-mw_fluence_dist} should be handled using
Equation~(\ref{E-mw_fluence-prob}).

The results are close to those of \inlinecite{Trottet2015}. All of
the scatter plots show a similar direct tendency with a scatter of
the same order. Correlations in Figure~\ref{F-cor4}a and
\ref{F-cor4}b are considerably lower for 22 west events with
$\lambda > 20^{\circ}$ than for all events because of a large
contribution from the four west abundant mM events. The higher
correlation of the proton fluence, $\Phi_\mathrm{100}$, with SXR
fluence, $\Phi_\mathrm{SXR}$, than with the peak SXR flux,
$I_\mathrm{SXR}$, is consistent with the significance of both the
intensity and duration of the acceleration process. On the other
hand, the contribution from the BFS is not excluded.

Figure~\ref{F-cor4}a additionally indicates that the 4 November
1997 event (SOL1997-11-04T05:58, 51 in Table~\ref{T-data_table})
probably belongs to the abundant events, too. This event was
associated with a short-duration (3~min) microwave burst up to
1000~sfu and a relatively slow CME (785~km~s$^{-1}$), but its
proton fluence was atypically high relative to the events with
comparable microwave fluences. In its SXR peak flux of X2.1, this
event is not atypical. By its SXR fluence and the CME speed, this
event resides in the upper part of the main cloud of points.

It is reasonable to assume that in the proton-abundant events,
depending on their location relative to the main cloud of points in
Figure~\ref{F-cor4}, the contribution of shock-accelerated protons
dominated. It is also possible that some additional factors were
implicated, especially for the event 2000-11-08, which stands apart
in Figure~\ref{F-cor4}d by its abundant proton fluence, while the
CME speed of 1738~km~s$^{-1}$ is insufficient to fit within the
cloud of points.

Now we consider the remaining events, excluding the abundant mM
events. For convenience we have plotted in Figures \ref{F-cor4}a
and \ref{F-cor4}d the ellipses enclosing all of the non-abundant
events. The ellipticity is known to visually characterize the
correlation coefficient. The points inside the dotted ellipse in
Figure~\ref{F-cor4}a are obviously least scattered with respect to
the parameters of the SXR emission and CME speed in other panels.
Thus, the BFS measure referring to the SXR emission used by
\inlinecite{Kahler1982} cannot account for the high correlation
between the microwave and proton fluences. This close correlation
persists over three orders of magnitude for $\Phi_\mathrm{35}$ and
five orders of magnitude for $\Phi_\mathrm{100}$. It would be
surprising to have this conspicuous correspondence as an
insignificant secondary effect due to BFS.

The dashed ellipse in Figure~\ref{F-cor4}d characterizes the
importance of the shock-acceleration according to
\inlinecite{Trottet2015}. While almost all of the abundant events
fall within the ellipse, the scatter here is obviously larger than
in Figure~\ref{F-cor4}a. The close correlation between
$\Phi_\mathrm{35}$ and $\Phi_\mathrm{100}$ cannot be a result of
the scattered correlation between $V_\mathrm{CME}$ and
$\Phi_\mathrm{100}$ due to the interdependence of the analyzed
parameters. Thus, it cannot be caused by the BFS. For reliability,
statistical characteristics of these relations are examined
quantitatively in the next section.

The range of the CME speeds is one order of magnitude, being limited
from below by about 400~km~s$^{-1}$, suggesting a lower limit
required for CME to drive a bow shock. On the other hand, CMEs spend
large energy to overcome gravity \cite{UralovGrechnevHudson2005}.
The gravity escape velocity at the inner boundary of the LASCO/C2
field of view, $r_{(\mathrm{C2})} = 2R_\odot$, is
$\sqrt{2GM_\odot/r_{(\mathrm{C2})}} \approx 440$~km~s$^{-1}$ ($G$
the gravity constant, $R_\odot$ and $M_\odot$ the radius and mass of
the Sun). Slower CMEs, whose propelling forces cease at lesser
heights (mainly from small sources), would not stretch closed
structures enough to enable efficient escape of trapped
flare-accelerated particles, or even fall back without any
appearance in the LASCO/C2 field of view. The majority of the
escaping slower CMEs is probably due to eruptions of large quiescent
filaments gradually accelerating up to large distances. SPEs are not
expected from such CMEs, which are too slow to produce shock waves
and are non-flare-related. The lower limit for the speeds of
SPE-related CMEs of about 400~km~s$^{-1}$ is expected in any case.

\subsection{Analysis of the Correlations}
 \label{S-partial_correlations}

To exclude secondary correlations between parameters, which are
not related physically, we use partial correlation coefficients,
following the approach of \inlinecite{Trottet2015}. Unlike the
classical Pearson correlation coefficients, the partial
correlation coefficients reveal the own contribution from each of
the parameters, suppressing the interdependence between them. The
partial correlation coefficient, $\rho_\mathrm{j}(\mathrm{x_{j}},
\mathrm{y})$, between the analyzed parameter, $\mathrm{x_{j}}$,
and the dependent random variable, $\mathrm{y}$, is calculated as
the usual Pearson correlation coefficient between $\mathrm{x_{j}}$
and the difference $(\mathrm{y} - \mathrm{Y_{j}})$, where
$\mathrm{Y_{j}}$ is the best linear fit of $\mathrm{y}$,
calculated from other parameters. The linear regression is used as
the best fit $\mathrm{Y_{j}}$:
$$\mathrm{Y_{j}} = \mathrm{C} + \sum\limits_{i \neq j}{x_i}.$$
\noindent The partial correlation coefficient can be considerably
less than the Pearson coefficient, but cannot exceed it.

Table~\ref{T-table_cor_all_par} presents the Pearson and partial
correlation coefficients between the analyzed parameters for all
28 events (columns 2 and 3) and for 23 events, excluding the
abundant mM events (columns 4 and 5). The correlation coefficients
were calculated for both the actual proton fluences (columns 2 and
4) and for the longitude-corrected ones (columns 3 and 5). The
Pearson correlation coefficients in the four upper rows of column
(3) correspond to Figure~\ref{F-cor4}. For comparison, column (6)
lists the results obtained by \inlinecite{Trottet2015} for peak
proton fluxes, $J_{15}$, of a lower energy range of 15--40 MeV.

 \begin{table} 
 \caption{Correlations between parameters of the solar eruptive activity
and SPE fluences in 28 events (1996--2014) in comparison with
results of Trottet \textit{et al.} (2015)}
 \label{T-table_cor_all_par}
 \begin{tabular}{lrrrrr}
 \hline

\multicolumn{1}{c}{Correlation coefficients} & \multicolumn{4}{c}{$\log_{10} \Phi_{100}$} & \multicolumn{1}{c}{$\log_{10} J_{15}$} \\
 & \multicolumn{2}{c}{All}  & \multicolumn{2}{c}{Without abundant} & \multicolumn{1}{c}{Results of} \\
 & \multicolumn{2}{c}{events (28)}  & \multicolumn{2}{c}{events (23)} & \multicolumn{1}{c}{Trottet \textit{et al.}}\\
 & \multicolumn{1}{c}{Actual} & \multicolumn{1}{c}{Corrected}  & \multicolumn{1}{c}{Actual} & \multicolumn{1}{c}{Corrected} &  \multicolumn{1}{c}{(2015)}\\

 \multicolumn{1}{c}{(1)} & \multicolumn{1}{c}{(2)} & \multicolumn{1}{c}{(3)} &
 \multicolumn{1}{c}{(4)} & \multicolumn{1}{c}{(5)} & \multicolumn{1}{c}{(6)} \\

 \hline

Pearson cor. coef. &  &  &  & & \\

$\log_{10} \Phi_{35}$ & 0.58 & 0.63 & 0.89 & 0.90 & 0.67 \\
$\log_{10} I_\mathrm{SXR}$ & 0.47 & 0.52 & 0.73 & 0.74 & 0.54 \\
$\log_{10} \Phi_\mathrm{SXR}$ & 0.73 & 0.73 & 0.82 & 0.79 & 0.76 \\
$\log_{10} \mathrm{V}_\mathrm{CME}$ & 0.71 & 0.71 & 0.77 & 0.75 & 0.67 \\

Partial cor. coef. &  &  & &  & \\

$\log_{10} \Phi_{35}$ & $-0.02$ & 0.09 & 0.59 & 0.67 & $-0.10$ \\
$\log_{10} I_\mathrm{SXR}$ & $-0.37$ & $-0.25$ & $-0.16$ & 0.09 & 0.06 \\
$\log_{10} \Phi_\mathrm{SXR}$ & 0.57 & 0.47 & 0.36 & 0.10 & 0.42 \\
$\log_{10} \mathrm{V}_\mathrm{CME}$ & 0.39 & 0.33 & 0.14 & 0.001 & 0.36 \\

\multicolumn{5}{c}{Without account of $\mathrm{V}_\mathrm{CME}$} & \\

$\log_{10} \Phi_{35}$ & 0.18 & 0.27 & 0.70 & 0.74 & \\
$\log_{10} I_\mathrm{SXR}$ & $-0.38$ & $-0.27$ & $-0.14$ & 0.10 & \\
$\log_{10} \Phi_\mathrm{SXR}$ & 0.64 & 0.55 & 0.35 & 0.10 & \\

 \hline
 \end{tabular}
 \end{table}

As \inlinecite{Trottet2015} analyzed only the west events (without a
longitude correction), to which both flare-related and shock-related
contributions were possible, their results in column (6) should be
compared with column (3). The Pearson correlation coefficients in
these columns are close to each other. The partial correlation
coefficients in columns (3) and (6) for $\Phi_\mathrm{35}$ and
$I_\mathrm{SXR}$ are more distinct, while the overall conclusion of
\inlinecite{Trottet2015} is confirmed. The correlations between the
SPEs and either the total microwave fluences or the SXR peak fluxes
for all 28 events are insignificant. The correlations of high-energy
proton fluences with the start-to-peak SXR fluences and the CME
speeds are significant. The partial correlation coefficients of
$\Phi_{100}$ with $\Phi_\mathrm{SXR}$ and with $V_\mathrm{CME}$ are
close to each other. Thus, our results for 28 events agree with
those of \inlinecite{Trottet2015}.

Columns (4) and (5) present the results for the same set of
events, except for the five abundant events. The Pearson
correlation coefficients between the proton fluences and all
independent parameters considerably increase, depending on the
longitude correction weakly. The partial correlation coefficients
increase sharply for $\Phi_\mathrm{35}$ and considerably reduce
for $V_\mathrm{CME}$. The influence of the longitude correction is
strong here. The dependence on the SXR emission becomes weak,
probably due to an indirect correlation via $\Phi_\mathrm{35}$.
These circumstances apparently confirm the predominant flare
origin of high-energy protons in the 23 events (columns 4 and 5),
and the major contribution from the shock-acceleration in the five
abundant events. The longitudinal correction sharply decreases the
significance of $\Phi_\mathrm{SXR}$, probably, due to its
interdependence with $\Phi_\mathrm{35}$.

We are not aware of the CME speeds for 15 proton events of
1990--1992 and event 54 (1998-11-22). Therefore, a rigorous analysis
of the partial correlation coefficients with all of the analyzed
parameters for the complete set of the 40 SPEs from 1990 to 2015 is
not possible. We will try to approximately estimate the significance
of the contributions from different sources by calculating the
partial correlation coefficients with no account of $V_\mathrm{CME}$
for the considered sets of 28 and 23 events. They are shown in the
bottom three rows of Table~\ref{T-table_cor_all_par}. Then we
compare these values with similar results obtained for all 40
events.

The partial correlation coefficients with $I_\mathrm{SXR}$ in all
columns and with $\Phi_\mathrm{SXR}$ in columns (4) and (5) are
almost insensitive to the absence of $V_\mathrm{CME}$. The largest
increase show the partial correlation coefficients with
$\Phi_{35}$ in columns (2) and (3), possibly due to their
interdependence with $V_\mathrm{CME}$. Their increase is not as
large in the columns (4) and (5), where the influence of
$V_\mathrm{CME}$ is less.

The correlation coefficients for all 40 SPEs from our list,
excluding the five abundant events, are listed in
Table~\ref{T-table_cor_all_events}. The Pearson correlation
coefficients between the proton fluences and the other known
parameters are not much different from the values in columns (4)
and (5) of Table~\ref{T-table_cor_all_par}. The partial
correlation coefficients with $\Phi_\mathrm{35}$ slightly
decrease, remaining the largest ones. Similar to column (5),
$\Phi_\mathrm{SXR}$ turns out to be insignificant, while the
importance of $I_\mathrm{SXR}$ sharply increases. This fact is not
surprising, because, firstly, $\Phi_\mathrm{35}$ is related to the
total flare energy, while $I_\mathrm{SXR}$ is associated with its
maximum power. Both parameters can be important. Secondly, an
indirect correlation between the proton fluence and
$I_\mathrm{SXR}$ through the unknown $V_\mathrm{CME}$ is possible.
The microwave fluence is most significant anyway.

 \begin{table} 
 \caption{Correlations between parameters of flares and SPE fluences in
40 presumably flare-dominated events (1990--2014)}
 \label{T-table_cor_all_events}
 \begin{tabular}{lrr}
 \hline

 \multicolumn{1}{c}{} & \multicolumn{2}{c}{$\log_{10} \Phi_{100}$} \\
 & Actual & Corrected  \\
 \hline

Pearson correlation coefficients &  &   \\

$\log_{10} \Phi_{35}$ & 0.77 & 0.82  \\
$\log_{10} I_\mathrm{SXR}$ & 0.66 & 0.75 \\
$\log_{10} \Phi_\mathrm{SXR}$ & 0.66 & 0.74  \\

Partial correlation coefficients &  &   \\

$\log_{10} \Phi_{35}$ & 0.56 & 0.62  \\
$\log_{10} I_\mathrm{SXR}$ & 0.33 & 0.46 \\
$\log_{10} \Phi_\mathrm{SXR}$ & $-0.16$ & $-0.15$  \\

 \hline
 \end{tabular}
 \end{table}

The results lead to the following conclusions. i)~The quantitative
analysis of the correlation coefficients confirms the conclusions
of the preceding Section~\ref{S-relations}. ii)~The partial
correlation coefficients are sensitive to the analyzed set of
events and allow one identifying significant parameters, but do
not guarantee independence of other parameters. iii)~The
predominance of the flare contribution to high-energy proton
enhancements in the 40 events of 1990--2015 with $F_\mathrm{35}
\geq 10^3$~sfu seems to be undoubted, while the contribution from
shock-acceleration is not excluded in these events. iv)~Similarly,
the predominance of the shock-acceleration in the five abundant
events seems to be certain, while the flare contribution in these
events is not excluded, too.

In summary, the quantitative analysis confirms the apparent
outcome from Figures \ref{F-plot4}d and \ref{F-cor4}d. The
statistical predominance of the flare-related contribution to SPEs
after the bursts with $F_\mathrm{35} \geq 10^3$~sfu observed
during 25 years by NoRP is confirmed by the scatter plot in
Figure~\ref{F-plot4}d. It does not reveal conspicuous outliers,
except for the five presumably shock-dominated abundant events.
Some of them in the $V_\mathrm{CME} - \Phi_{100}$ scatter plot
(Figure~\ref{F-cor4}d) surpass in proton fluences their neighbors
in the cloud of the points. This circumstance indicates a possible
implication of some factors amplifying their proton productivity.

\section{Discussion and Conclusion}
 \label{S-discussion}

\subsection{Results of the Analysis}
 \label{S-results}

The dependence of the probability of a proton enhancement on the
duration of the microwave burst was analyzed, its empirical
quantitative description was proposed. Unlike the traditional
estimate of the flare duration from its SXR emission, we used the
duration of the microwave burst at 35 GHz,
$\Delta\mathrm{t_{35}}$. According to the Neupert effect
\cite{Neupert1968}, $\Delta\mathrm{t_{35}}$ should be close to the
duration of the rise phase in SXR. Therefore, the difference
between our estimates and the traditional way should not be large.

Clustering the events according to the durations of the 35~GHz
bursts, expected for the categories of `impulsive' and `gradual'
events, was not revealed for the proton enhancements of $>
100$~MeV. This result might be due to a usual displacement of the
microwave turnover frequency below late in long-duration events,
decreasing the flux density at 35 GHz. Considerable efforts were
applied previously to search for a criterion or index of the flare
`impulsiveness' (\textit{e.g.}, \opencite{Cliver1989}), but no
certain quantitative result was obtained. Two presumable
categories, differing in their particle composition and other
properties, were distinguished in their durations only
qualitatively. To clarify the situation, we analyzed the
correlations between all combinations of the peak fluxes and
fluences for the proton enhancements and the microwave bursts. For
the majority of the analyzed events, the highest correlation was
found between the proton fluences, on the one hand, and the
microwave and SXR fluences, on the other hand
(Figure~\ref{F-cor4}). In other words, the total number of
near-Earth protons is controlled by both the intensity of the
particle acceleration process and its duration. This circumstance
points at a correspondence between the durations of the proton
acceleration process and the flare that is obvious for the flare
origin of protons, but more difficult to understand for the proton
acceleration by shock waves. Thus, a probable reason for the
dependence of the number of high-energy protons on the duration of
an event is not a difference between the particle acceleration
mechanisms, but the duration of the acceleration process.

Some conceptions of the properties of the two different categories
of `impulsive' and `gradual' events might be probably due to the
traditional idea that impulsive events are associated with
confined flares and long-duration events (LDEs) are associated
with eruptive flares. However, two decades of SOHO/LASCO
observations have shown that this assumption was oversimplified.
Indeed, most confined flares are short, and most LDEs are
associated with CMEs. However, many impulsive flares are obviously
eruptive, \textit{e.g.}, events 19, 51, 55, 56, 58, 66--69, 90,
93, and 94 in Table~\ref{T-data_table}. They include GOES X-class
flares and halo CMEs. On the other hand, confined LDE flares are
known, \textit{e.g.}, some of the events in active region 12192 in
October 2014 (\opencite{Thalmann2015}; also, \textit{e.g.}, event
101 in Table~\ref{T-data_table}). These circumstances indicate
that the duration of an event is not a reliable indicator of a
dominating acceleration mechanism. In particular, one of the five
presumably shock-dominated abundant events (51 in
Table~\ref{T-data_table}) was an impulsive one.

Recently, \inlinecite{Reames2014} analyzed the SEP composition of
111 impulsive events with a high iron abundance and concluded that
the sources of their major part were associated with CMEs. It is
interesting to analyze a possible overlap of the events from their
list with the NoRP data which we considered. We found that the
sources of 39 events from that list fell within the NoRP
observation time. Four events are present in our
Table~\ref{T-data_table} (56, 59, 64 and 93); weak proton
enhancements above 100~MeV were observed after three of them. One
major proton event (2004-11-01, 05:50) was probably due to a
behind-the-limb source. The peak fluxes at 35 GHz did not exceed
100~sfu in 19 events. No bursts at 35 GHz were detectable in 16
events. Most of these events did not produce noticeable proton
enhancements even in the $> 10$~MeV range.

In this respect, a direct dependence of the probability of proton
enhancements on the peak intensity of microwave bursts at 35 GHz
seems to deserve attention. It is important for diagnostic
purposes. Possibly, such a dependence can be manifest at
frequencies below 35 GHz, for which round-the-clock observations
are more representative. The only event from the analyzed set,
with a 35 GHz peak flux of $10^{3}$~sfu, was probably
shock-dominated. These facts suggest a possible indication at a
dominating flare-related source of high-energy SPE, if
$F_\mathrm{35} > 10^{3}$~sfu, or a prevailing shock-related
source, if $F_\mathrm{35} < 10^{3}$~sfu.

\subsection{Are the Two Alternative Concepts Really Incompatible?}
 \label{S-contesting_concepts}

As mentioned, the statistical domination of flare-related
acceleration of high-energy proton enhancements in most of the
events, which we analyzed, does not exclude a contribution from
shock-related acceleration in these events. It is supported, for
example, by the analyses of the SEP composition (\textit{e.g.},
\citeauthor{Reames2009}, \citeyear{Reames2009, Reames2013}, and
many others). Recent observational studies (\opencite{Qiu2007},
\opencite{Temmer2010}, \citeauthor{Grechnev2011}
\citeyear{Grechnev2011, Grechnev2013a}) have revealed a closer
association between solar eruptions, flares, shock waves and CMEs,
than previously assumed. The shock waves initially appear early in
the low corona, during the rise phase of a hard X-ray and
microwave burst. Particle acceleration by flare processes and
shock waves can occur nearly concurrently, and therefore it is
hardly possible to recognize their origin from the analysis of
temporal relations or velocity dispersion. On the other hand, the
account of the early appearance of shock waves at low altitudes
can be helpful in the studies of the SEP acceleration by shock
waves.

Conclusions about the origin of near-Earth proton enhancements made
on the basis of oversimplified old hypotheses without confronting
with recent observations might be inadequate. Taking into account
the results of our analysis and those of \inlinecite{Trottet2015},
one can expect that shock-acceleration is responsible for the bulk
of protons and ions accelerated to low to moderate energies. On the
other hand, the major role of shock waves in the acceleration of GLE
particles, which represent the SPE category with a hardest spectrum,
looks questionable in events associated with power flares. Indeed,
apparently shock-dominated non-flare-related SPEs are characterized
by soft spectra (see, \textit{e.g.}, \opencite{Chertok2009};
\opencite{Gopalswamy2015}). The hardness of the proton spectra in
GLE-related events is confirmed by our data set. All of such events
(marked with a superscript `$^{1}$' in column (13) of
Table~\ref{T-data_table}) had proton indices $\delta_\mathrm{p} <
1.5$. Protons with energies above 100 MeV are sometimes observed in
non-flare-related SPEs, but their percentage is less than in
flare-related events. For example, a detailed analysis of the origin
of SPE in the extreme 20 January 2005 event responsible for GLE69
led \inlinecite{Grechnev2008} and \inlinecite{Klein2014} to the
conclusions about the flare source of SPEs. Similarly, the analysis
by \inlinecite{Grechnev2013a} of the 13 December 2006 event
responsible for GLE70 revealed inconsistency of previous arguments
in favor of its exceptional shock-related source. It is possible
that in exceptional non-flare-related events shock-accelerated
proton fluxes are sufficient to produce a GLE under favorable
conditions \cite{Cliver2006}. However, it is difficult to expect
that if a powerful flare occurs, then shock-accelerated protons
provide the main contribution to the GLE, relative to the
flare-related contribution dominating at high energies. Note that we
analyzed the GOES integral proton channel above 100 MeV, although
particles of much higher energies, $\gsim 1$~GeV, are responsible
for GLEs.

These considerations seem to be opposed by a recent study of
\inlinecite{Thakur2014}, where the authors came to a conclusion
about exceptional shock-related origin of the 6 January 2014 GLE.
It was produced by a behind-the-limb SOL2014-01-06 event in active
region 11936 (Table~\ref{T-table_01-2014}), where the STEREO
telescopes recorded a powerful flare with an estimated GOES
importance of about X2 \cite{Chertok2015}. Proceeding from the
remoteness of the flare region from the well-connected longitudes
(like event 110 in Table~\ref{T-data_table} associated with GLE61
on 18 April 2001), \inlinecite{Thakur2014} stated that this event,
as well as other GLEs from behind-the-limb sources, posed a
challenge to the flare acceleration mechanism for GLE particles.
However, as the facts and speculations in the preceding paragraph
show, the shock-related origin of some GLEs does not contradict
the major flare contribution to the others.

 \begin{table} 
 \caption{Comparison of the proton events on 6 and 7 January 2014}
 \label{T-table_01-2014}
 \begin{tabular}{cccrrrr}
 \hline

\multicolumn{1}{c}{Event} & \multicolumn{1}{c}{Position} & \multicolumn{1}{c}{GOES class} &
 \multicolumn{1}{c}{CME speed} & \multicolumn{1}{c}{$J_{100}$} & \multicolumn{1}{c}{$J_{10}$} &
 \multicolumn{1}{c}{$\delta_\mathrm{p}$} \\
 &  &  & \multicolumn{1}{c}{[km~s$^{-1}$]} & \multicolumn{1}{c}{[pfu]} & \multicolumn{1}{c}{[pfu]} &  \\

 \hline

 SOL2014-01-06T07:50 & S15W113 & $\approx$ X2 & 1402 & 4 & 40 &
1.00
\\

 SOL2014-01-07T18:32 & S11W11 & X1.2 & 1830 & 4 & 900 & 2.35
\\

 \hline
 \end{tabular}
 \end{table}

Assuming a direct escape of flare-accelerated protons into the
interplanetary space from the active region core in the low corona,
\inlinecite{Thakur2014} pointed out that the flare-accelerated
particles would need to interact with the CME flux-rope to reach the
well-connected field lines, and thus their scattering would not
allow the high anisotropy typical of the beginning of the GLEs. It
is difficult to agree with this argument, because the major cause of
the anisotropy of GLE particles is their transport in the
interplanetary space. There are some other complicating factors such
as the perpendicular diffusion that is difficult to take into
account in simple considerations. A surprising example presents an
SPE caused by the 1 September 2014 event behind the east limb, when
the rise phase during half a day was dominated by $> 100$~MeV
protons. Furthermore, there is a possibility of trapping of
accelerated protons in the CME flux-rope trap (similar to electrons
responsible for type IV radio bursts) and their confinement until
reconnection of the flux rope with an open magnetic structure such a
coronal hole or streamer that allows the trapped particles access to
the interplanetary space (see, \textit{e.g.}, \opencite{Masson2012};
\opencite{Grechnev2013b}). In such a case, the escape conditions for
protons accelerated in a flare and at the shock front ahead of a CME
are practically the same.

It is useful to compare the 6 January event with another, which
occurred on the next day, 7 January 2014
(Table~\ref{T-table_01-2014}), in active region 11944 on the
Earth-facing solar side, and also produced an SPE, but not a GLE.
The peak flux of $> 100$~MeV protons was the same as on 6 January,
while the lower-energy SPE was stronger and longer. The 7 January
CME was considerably faster than the 6 January CME. The peak flux
of the gyrosynchrotron emission of 2500~sfu occurred on 7 January
at about 5 GHz. All of the GLE-related events from our sample had
a higher turnover frequency \cite{Grechnev2013b}; parameters of
the microwave emission on 6 January are unknown. The relation
between the parameters of the two events is not surprising, if the
softer shock-related SPE component dominating lower energies was
stronger on 7 January, while the harder flare-related component
dominating higher energies was stronger on 6 January. Otherwise,
the relation seems to be challenging. For all of the listed
reasons, the arguments against the flare-related source of the 6
January 2014 GLE are not convincing.

The existence of the two concurrent different sources of
accelerated protons has been argued previously in several studies
mentioned in Section~\ref{S-introduction}. This alternative to the
single-source hypothesis invoked by \inlinecite{Kahler1982} can
also be checked by comparing the peak-size distributions of SPEs
and all flares. Analyzing the differential distribution functions
\textit{vs.} energy, \inlinecite{Hudson1978} concluded that the
proton production was more efficient in more energetic flares.
Three decades later, \inlinecite{Belov2007} have analyzed detailed
distribution based on much richer data. They demonstrated that the
slope of the distribution of SPE-related flares was flatter than
that of all flares at their low to moderate GOES importance
(\textit{i.e.}, the SPE productivity of weaker events was less
dependent on the SXR peak flux). For bigger flares, the slope of
SPE-related flares approached that of all flares. These
circumstances also confirm the existence of the two sources of
SPEs; one, shock-related, dominates in events with weaker flares,
and the second, flare-related, dominates in stronger flares. Note
that the analyses of the peak-size distributions did not consider
the event duration, whose role we discussed
(\inlinecite{Hudson1978} also admitted this possibility).

All of the listed facts indicate that the role of the BFS was most
likely overestimated by \inlinecite{Kahler1982}. He stated a
higher correlation between the 8.8 and 15.4 GHz fluences and SPE
peak fluxes than the BFS could provide, but did not consider this
correlation to be important, having not found any correspondence
between the spectral parameters of SPEs and microwaves (similar
conclusion was made about hard X-rays). Some aspects of this
correspondence have been revealed later. \inlinecite{Chertok2009}
demonstrated statistical correspondence between
$\delta_\mathrm{p}$ and spectral parameters of microwave bursts.
\inlinecite{Grechnev2013b} showed that the SPEs produced in events
with $F_{35} > 10^3$~sfu were harder than those after weaker
bursts. The results of \inlinecite{Kahler1982} might be determined
by the limitations of the microwave data used in his analysis.
Most of the 50 events he analyzed had peak microwave fluxes from
$10^2$ to $10^4$~sfu in the whole frequency range; two only were
stronger. When referred to 35 GHz, the majority of these events
fall into the mM category, suggestive of prevailing
shock-acceleration of SPEs, as we showed. Thus, the extension of
our results to SPEs associated with weaker microwave bursts should
mainly correspond to the results of \inlinecite{Kahler1982}.

\subsection{Concluding Remarks}
 \label{S-conclusion}

Our analysis has not revealed a separation of the analyzed data
set at 35~GHz according to their durations into the clusters of
`impulsive' and `gradual' events. Relations have been established
between the intensities and durations of microwave bursts, on the
one hand, and the probability of near-Earth proton enhancements
with energies $> 100$~MeV, on the other hand. Most likely, the
causes of these dependencies are related to propagation effects of
protons from their solar sources to the Earth and the limited
sensitivity of the detectors. This circumstance suggests the
possibility that protons are accelerated to high energies in all
flares accompanied by sufficiently strong bursts at 35 GHz,
\textit{i.e.}, in all cases, when acceleration of a large number
of electrons to relativistic energies occurs. This indication
corresponds to the conclusions of \inlinecite{Livshits2004} about
the simultaneous acceleration of both electrons and protons.

Our results are consistent with the main conclusion of
\inlinecite{Trottet2015} and confirm their suggestion about the
domination of the flare acceleration for high-energy protons. For
the majority of the analyzed events, we found a direct dependence
with a high correlation between the parameters of the flare and
proton fluences $> 100$~MeV. Comparable correlations between the
proton fluences with start-to-peak SXR fluences and microwave
emission shows that both these parameters characterizing solar
flares can be used for the diagnostics of proton enhancements, and
their importance is not diminished by the Big Flare Syndrome
hypothesis. Comparison of Figures \ref{F-long-cor}a,
\ref{F-long-cor}b, \ref{F-plot4}d, and
Table~\ref{T-table_cor_all_events} demonstrates that finding and
accounting the factors, which affect the quantitative parameters of
near-Earth proton enhancements, allow one to considerably reduce the
uncertainty of their expected values that is evaluated by a
conspicuous increase of the correlation coefficients. Perhaps some
other affecting factors exist, whose account would additionally
reduce the scatter. For example, by using a combination of the peak
flux, effective duration, and the turnover frequency of the
microwave bursts, \inlinecite{IsaevaMelnikovTsvetkov2010} reached
their considerably higher correlation with SPE parameters.

A detailed analysis of recent observational data promises a
substantial progress in understanding the sources of near-Earth
proton enhancements and their prompt forecast. It seems attractive
to analyze the events, in which the contribution from only one of
the two competing sources of accelerated protons is most probable.
Note, however, that, according to recent observational studies,
shock waves develop in the low corona during flares and the early
formation of CMEs. This update can help in studies of particle
acceleration by shock waves, but it makes recognizing the sources
of SPEs still more difficult. Most likely, the events with
exceptional flare-acceleration do not exist, because shock waves
develop even in eruptive events without detectable microwave
bursts, while the escape of accelerated protons from confined
flares is hampered. On the other hand, SPEs without powerful
flares but with strong shock waves are known. Case studies of
selected events of such a kind might shed more light on one of the
two sources of SPEs. The results of these studies would provide
the guidelines for future statistic analysis.

\begin{acks}
We thank A.V.~Belov, B.Yu.~Yushkov, V.G.~Kurt, M.A.~Livshits,
H.~Nakajima, K.-L.~Klein, N.V.~Nitta, V.E.~Sdobnov, V.F.~Melnikov,
V.V.~Zharkova, and S.S.~Kalashnikov for useful discussions and
assistance. We thank an anonymous referee for valuable comments. We
are grateful to the instrumental teams operating Nobeyama Radio
Polarimeters and GOES satellites for the data used here. The CME
Catalog used in this article is generated and maintained at the CDAW
Data Center by NASA and the Catholic University of America in
cooperation with the Naval Research Laboratory. SOHO is a project of
international cooperation between ESA and NASA. The authors thank
the CDAW Team for the data. This study was supported by the Russian
Foundation of Basic Research under grants 14-02-00367, 15-02-10036,
and 15-02-01089, by the Russian Science Foundation under grant
16-12-00019, and the Program of basic scientific research of RAS No.
II.16.1.6. N.M. was partly supported by the Marie Curie
PIRSES-GA-2011-295272 RadioSun project.

\end{acks}

\end{article}

\end{document}